\documentclass[letter]{aa} 

\usepackage{graphicx}
\usepackage{txfonts}
\usepackage{hyperref}

\begin{document}

   \title{Rotation-induced granular motion on the secondary component of binary asteroids: Application to the DART impact on Dimorphos}
   \titlerunning{Rotation-induced granular motion on the secondary component of binary asteroids}

   \author{H. F. Agrusa\inst{1}\thanks{hagrusa@astro.umd.edu}
          \and
          R. Ballouz\inst{2}
          \and
          A. J. Meyer\inst{3}
          \and
          E. Tasev\inst{4}
          \and
          G. Noiset\inst{4}
          \and
          \"{O}. Karatekin\inst{4}
          \and
          P. Michel\inst{5}
          \and
          D. C. Richardson\inst{1}
          \and
          M. Hirabayashi\inst{5}
          }

   \institute{
            Department of Astronomy, University of Maryland, College Park, MD, 20742, USA
         \and
            Johns Hopkins University Applied Physics Laboratory, Laurel, MD, 20723, USA
         \and
            Smead Department of Aerospace Engineering Sciences, University of Colorado Boulder, Boulder, CO 80303, USA
         \and
            Royal Observatory of Belgium, 3 Avenue Circulaire, 1180 Brussels, Belgium
         \and
            Universit\'e C\^ote d’Azur, Observatoire de la C\^ote d’Azur, CNRS, Laboratoire Lagrange, Nice, France
         \and
            Department of Aerospace Engineering, Department of Geosciences, Auburn University, Auburn, AL, USA       
         }
   \date{Received June 30, 2022; accepted July 25, 2022}

 
  \abstract
   {NASA's Double Asteroid Redirection Test (DART) mission will kinetically impact Dimorphos, the secondary component of the Didymos binary asteroid system, which will excite Dimorphos's dynamical state and lead to significant libration about the synchronous state and possibly chaotic non-principal axis rotation. Although this particular outcome is human caused, many other secondary components of binary systems are also prone to such exotic spin states.}
   {For a satellite in an excited spin state, the time-varying tidal and rotational environment can lead to significant surface accelerations. Depending on the circumstances, this mechanism may drive granular motion on the surface of the secondary.}
   {We modeled the dynamical evolution of a Didymos-like binary asteroid system using a fully coupled, three-dimensional simulation code. Then, we computed the time-varying gravitational and rotational accelerations felt over the entire surface resulting from the secondary's perturbed dynamical state.}
   {We find that an excited spin and orbit can induce large changes in the effective surface slope, potentially triggering granular motion and surface refreshment. However, for the case of the DART impact, this effect is highly dependent on many unknowns, such as Dimorphos's detailed shape, bulk density, surface geology, and the momentum transferred. Aside from the Didymos system and the DART mission, this effect also has important implications for binary systems in general.}
   {}

   \maketitle
%

\vspace{-5pt}
\section{Introduction}
On September 26, 2022, NASA's Double Asteroid Redirection Test (DART) mission will kinetically deflect Dimorphos, the smaller component of the binary asteroid 65803 Didymos, as a planetary defense demonstration test \citep{Rivkin2021}. Prior to the impact, DART will deploy the Light Italian CubeSat for Imaging of Asteroids (LICIACube), which will fly by the system to image the initial phase of the cratering process as well as improve Dimorphos's shape determination \citep{Dotto2021,Cheng2022}. Following the impact, the change in the mutual orbit period will be measured via ground-based observations and used to infer the momentum enhancement factor, commonly referred to as $\beta$ \citep{Rivkin2021}. Due to the contribution of ejecta that exceeds the escape speed, $\beta$ is expected to exceed 1. Four years after DART, the European Space Agency's Hera mission will rendezvous with Didymos to characterize the physical, dynamical, and compositional properties of the system. Hera will also measure in detail the effects of the DART impact, including the crater's properties and the mass of Dimorphos, allowing for a more precise determination of $\beta$ \citep{Michel2022}. 

In addition to abruptly reducing the binary semimajor axis and orbit period, the impact will also change the eccentricity and inclination \citep{Cheng2016}. Due to a high degree of spin-orbit coupling, the dynamical evolution of Dimorphos strongly depends on the initial conditions at the time of impact and the body's shape, which are currently unknown \citep{Agrusa2020}. Depending on $\beta$ and Dimorphos's shape, it is possible that Dimorphos may enter a chaotic rotation state following the DART impact \citep{Agrusa2021,Richardson2022}. Furthermore, numerical simulations that treat Dimorphos as a rubble pile indicate that boulders may move on the surface, depending on Dimorphos's spin state, bulk shape, and material properties \citep{Agrusa2022a}. In this study, we take a closer look at the possibility of post-impact surface motion on Dimorphos as a function of its complex spin and orbital environment. 

Observational evidence and theoretical arguments both indicate that chaotic rotation is not uncommon for secondaries in tight binary systems \citep{Pravec2016,Cuk2021, Seligman2021, Quillen2022a}, and it is plausible that many synchronous secondaries have undergone some level of chaotic rotation in their past or during their formation \citep{Wisdom1987b,Jacobson2011a,Davis2020b}. Therefore, the methods and results presented here are also broadly applicable to the general binary asteroid population. 

\vspace{-5pt}
\section{Methods}
Focusing on the DART impact, we first ran a simulation to capture the system's dynamics, from which the local slopes can be computed, in an approach analogous to previous studies of dynamically triggered regolith motion \citep{Yu2014,Ballouz2019}. In order to capture the coupled spin and orbital motion of the secondary, we used the General Use Binary Asteroid Simulator (\textsc{gubas}), an efficient rigid full two-body problem (F2BP) code \citep{Davis2020a,Davis2021}. \textsc{gubas} has been benchmarked against other F2BP simulation codes and has been used extensively to study the dynamics of Didymos and other binary systems \citep{Agrusa2020, Davis2020b, Meyer2021a,Meyer2021b}. In accordance with previous studies, the \textsc{gubas} simulations expand the gravitational potential of the polyhedral shape models to degree and order 4 to adequately capture their irregular gravity fields. All simulations presented herein were run for 1 yr of integration time.

\begin{table}
    \caption{Selected physical and dynamical parameters used for the simulated Didymos system, consistent with the current best estimates \citep{Rivkin2021}. The body diameters are the volume-equivalent spherical diameters. A synchronous spin state for Dimorphos is {assumed}, and we refer the reader to \cite{Richardson2022} for further discussion on this assumption.}
    \label{tab:params}
    \centering 
    \begin{tabular}{ll}
        \hline \hline
        Parameter & Value \\
        \hline
        Primary bulk density ($\rho_\text{P}$) & $2.2$ g cm$^{-3}$ \\
        Secondary bulk densities ($\rho_\text{S}$) & $[1.85, 2.20, 2.55]$ g cm$^{-3}$\\
        Primary mass ($M_\text{P}$) & $5.47\times10^{11}$ kg\\
        Secondary masses ($M_\text{S}$) & $[4.20, 4.99, 5.78]\times10^9$ kg\\
        Primary Diameter ($D_\text{P}$) & $780$ m \\
        Secondary Diameter ($D_\text{S}$) & $164$ m \\        
        Initial body separation ($a_{\mathrm{orb}}$) & $1200$ m \\
        Initial Orbital Period ($P_{\mathrm{orb}}$) & $11.92$ h \\
        Primary Spin Period ($P_\text{P}$) & $2.26$ h \\
        Secondary Spin Period ($P_\text{S}$) & $11.92$ h\\
        Assumed DART Mass ($M_{\text{DART}}$) & 536 kg\\
        Assumed DART Speed ($v_{\text{DART}}$) & 6.15 km/s \\
        \hline
    \end{tabular}
\end{table}

  \begin{figure*}
      \centering
      \begin{minipage}[b]{0.3\hsize}
         \centering
         \includegraphics[clip, trim=2.25cm 5.22cm 1.9cm 4.8cm,width=\textwidth]{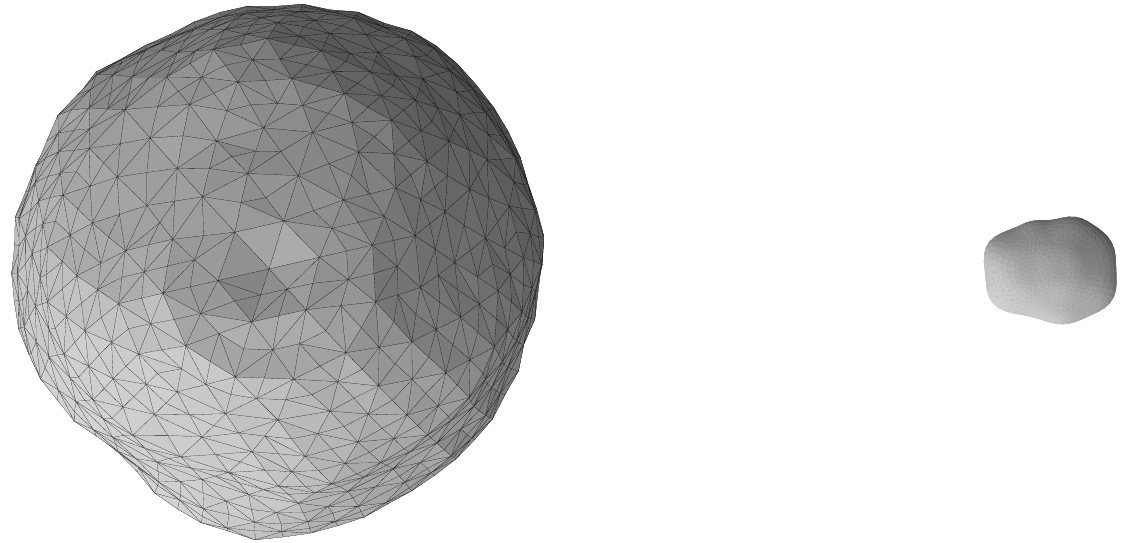}
         (a) 
         \includegraphics[clip, trim=2.25cm 5.1cm 1.75cm 4.6cm, width=\textwidth]{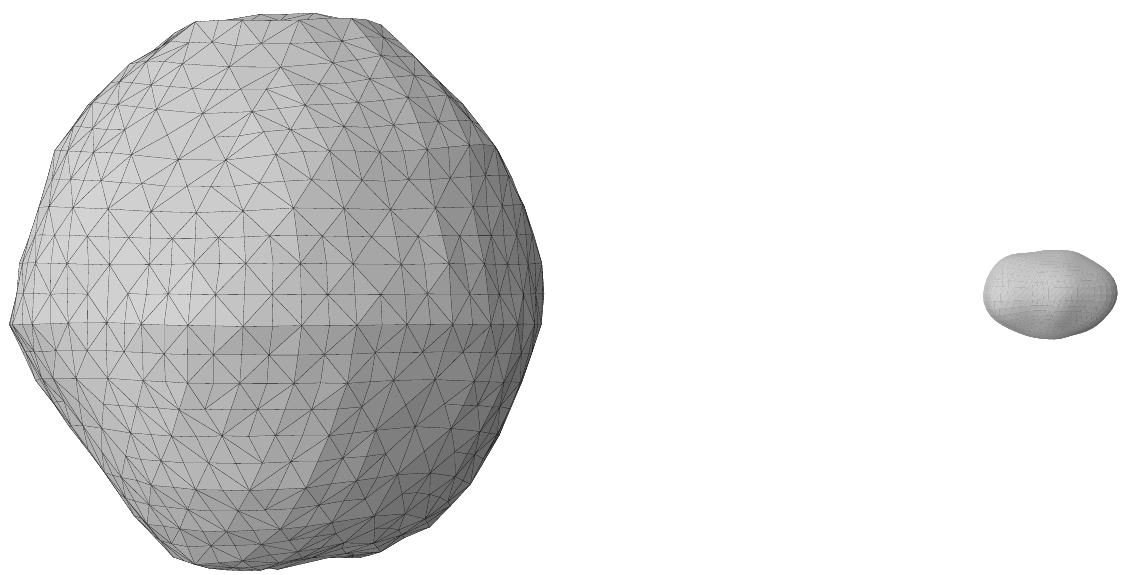}
         (b)
         \includegraphics[clip, trim=3cm 3.4cm 0.65cm 3.4cm, width=\textwidth]{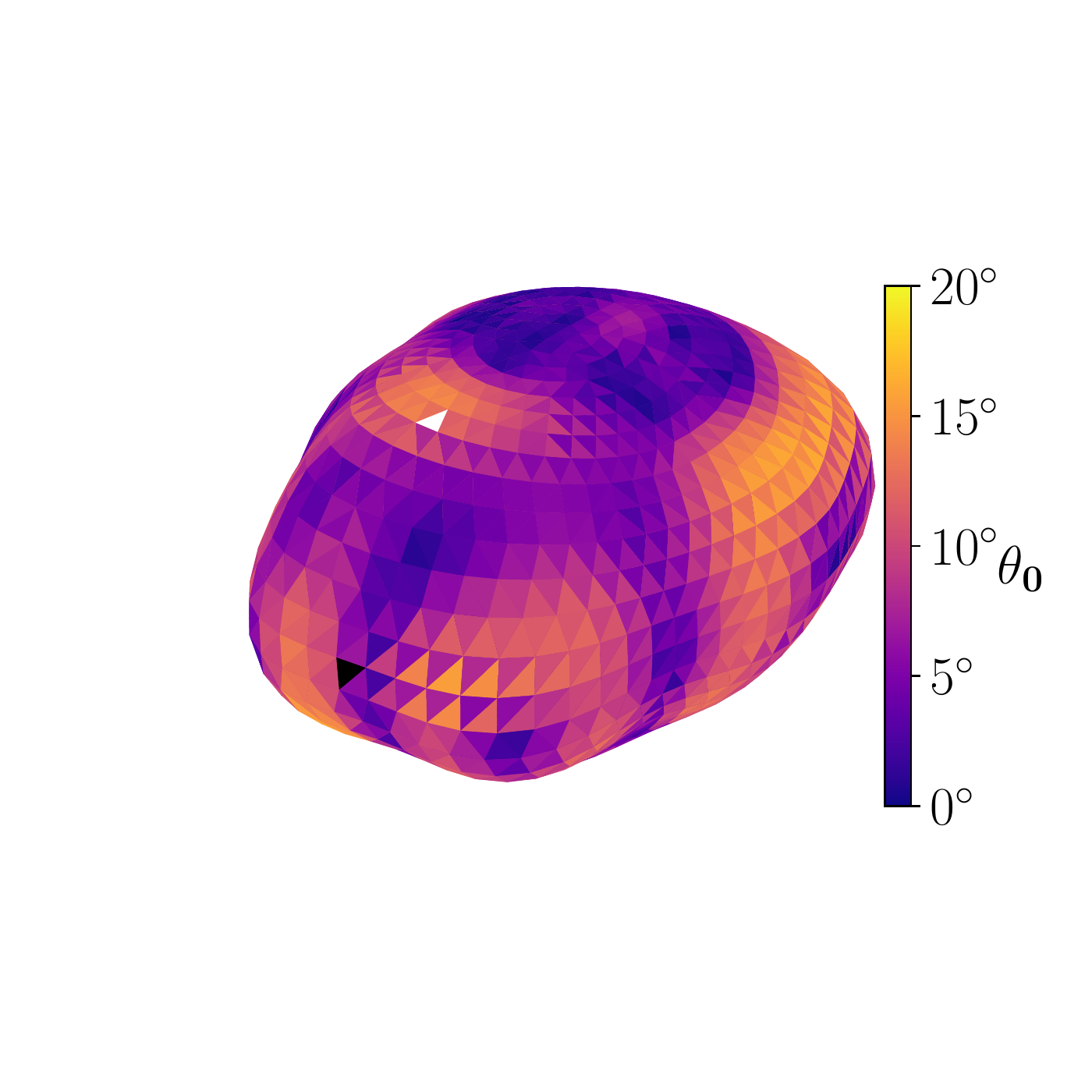}
         (c) 
      \end{minipage}      
      \begin{minipage}[b]{0.34\hsize}
         \centering
         \includegraphics[width=\textwidth]{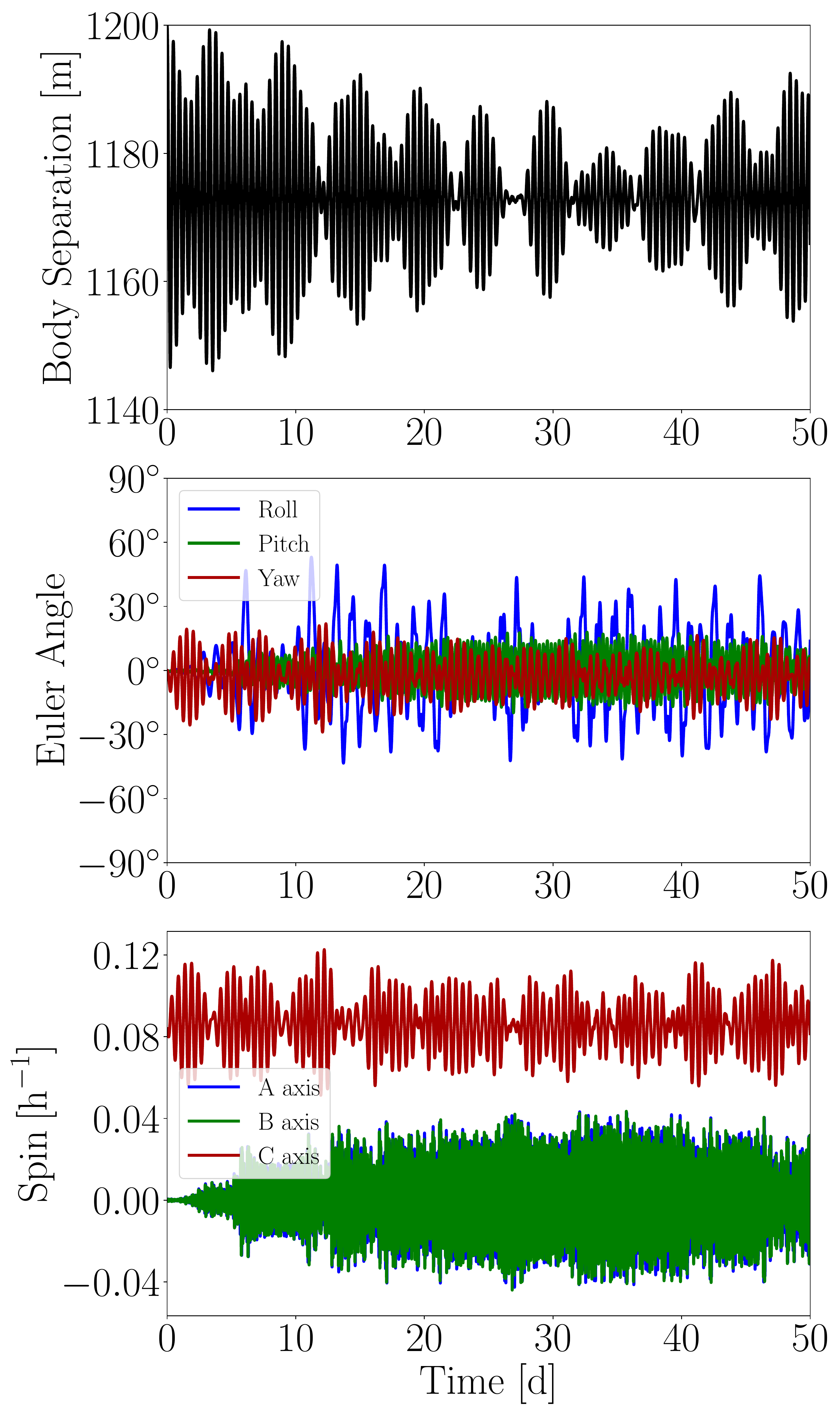}
         (d)
      \end{minipage}
      \begin{minipage}[b]{0.34\hsize}
         \centering
         \includegraphics[width=\textwidth]{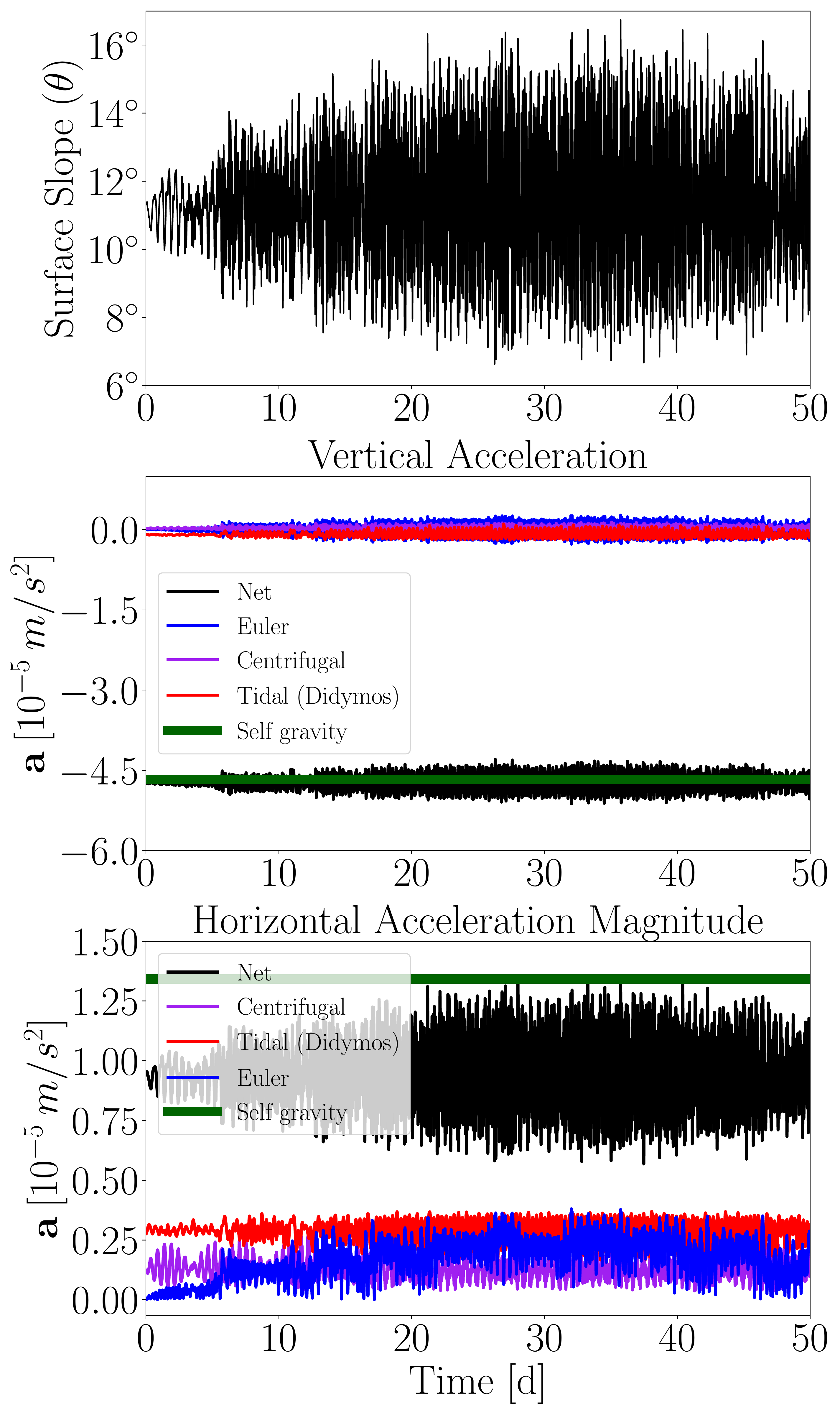}
         (e)
      \end{minipage}
      \caption{\label{fig:timeSeries} Surface slope evolution as a function of Dimorphos's dynamical evolution. (a) Top-down view of the ``Didymos-Squannit'' system. From this view, the spin and mutual orbit poles are pointing out of the page. (b) Side view. (c) Surface slopes for a Squannit-shaped Dimorphos with a bulk density of $\rho_\text{S}=2.2\text{ g cm}^{-3}$ in an idealized, relaxed dynamical state. The black facet corresponds to the sub-Didymos point (at zero libration amplitude) with a longitude and latitude of $\phi\approx\lambda\approx0^\circ$. The white facet has a longitude and latitude of  $(\phi, \lambda)\approx(0^\circ,45^\circ)$ and corresponds to the time-series plots in part (e). (d) Spin and orbital evolution for the Squannit-shaped Dimorphos when $\beta=3\, (e=0.023)$. The Euler angles are the 1-2-3 Euler angle set (roll-pitch-yaw) expressed in the rotating orbital frame, while the body spin rates are in the secondary's body-fixed frame. (e) Slope and surface accelerations on the white facet from part (c). The vertical accelerations point along the facet's surface normal and are generally dominated by self-gravity. The horizontal accelerations are expressed as magnitudes and point parallel to the surface. Initially, the Euler acceleration is relatively small and the tides are the dominant time-varying acceleration. After about 5 days, Dimorphos enters NPA rotation, and the Euler accelerations become comparable to both the tidal and centrifugal accelerations. We refer the reader to Appendix \ref{app:timeSeries} for an identical plot showing the full 365 d simulation.}
   \end{figure*}

\vspace{-5pt}
\subsection{Simulation setup}
In the F2BP simulations, the primary's gravity is modeled using Didymos's radar-derived polyhedral shape model \citep{Naidu2020}. Dimorphos's shape is still unknown, so we used the radar shape model for Squannit, the secondary component of the binary asteroid (66391) Moshup, scaled to the expected volume of Dimorphos. Squannit is arguably the best available analog for Dimorphos. Both the Didymos and Moshup systems are S types \citep{Binzel2004,Dunn2013} and have similar properties, including a fast-rotating primary with a raised equatorial ridge and a tidally locked secondary component on a tight, approximately circular orbit \citep{Scheeres2006b}.\footnote{There are no observations that show Dimorphos is spin locked, but circumstantial evidence indicates that this is likely. We refer the reader to \cite{Richardson2022} for a detailed discussion on this assumption.} Squannit is the only currently available secondary shape model for a near-Earth binary and contains ${\sim}2300$ facets \citep{Ostro2006}. Radar data tend to smooth and flatten surface features, making the surface slope analysis presented here somewhat conservative. When scaled to the dimensions of Dimorphos, Squannit's average facet has a surface area of ${\approx} 38 \text{ m}^2$. Schematics showing the shape models for the primary and secondary are shown in Fig.\ \ref{fig:timeSeries}(a-c). 

We focused this short study on the role of $\beta$ and Dimorphos's bulk density ($\rho_\text{S}$) as they play a significant role in determining the surface slope evolution of Dimorphos. The bulk density sets the mass and therefore the self-gravity of the body, which has a significant effect on the surface slope of a given shape model \citep{RichardsonJ2014,RichardsonJ2019}. For a fixed $\beta$, a smaller bulk density (i.e., lower mass) will result in a larger perturbation to the mutual orbit, which can lead to larger changes in surface slopes over time. We tested values of $\beta$ in the range $0\leq\beta\leq5$, in accordance with the best estimates from hydrodynamic simulations of the DART impact \citep{Raducan2022a,Stickle2022}. Based on light curve and radar observations, the Didymos system is expected to have a bulk density with $1\sigma$ uncertainties of $\rho{\approx}2.2 \pm 0.35\text{ g cm}^{-3}$ \citep{Naidu2020,Rivkin2021}. Assuming Dimorphos has a bulk density within this range, we tested values of $1.85, 2.2$, and $2.55\text{ g cm}^{-3}$. It should be noted that the reported uncertainties are for the bulk density of the {entire system}, which is of course dominated by the primary, and it is certainly possible for Dimorphos to have a bulk density outside of the range explored here (see the discussion on Dimorphos's density in \cite{Rivkin2021}). Table \ref{tab:params} provides the adopted physical and dynamical parameters for this study.

First, the binary was given dynamically relaxed initial conditions (i.e., a circular orbit with a synchronous secondary). Then, a change in velocity ($\Delta \vec{v}$) was applied to the secondary's instantaneous orbital velocity consistent with a head-on DART impact and a given selection for $\beta$ and $\rho_\text{S}$.\footnote{The DART impact will not be ideally head-on and centered, but recent work indicates that these effects should be negligible in terms of determining the system's bulk dynamical properties \citep{Richardson2022}.} This $\Delta \vec{v}$ reduces Dimorphos's velocity, causing the body to fall into a tighter, more eccentric orbit.\footnote{$\Delta \vec{v}$ is dependent on $\beta$, the impactor mass and velocity, as well as the secondary's mass. In a simplified scalar form, it can be written as $\Delta v = -\beta M_\text{DART}v_\text{DART}/M_\text{S}$, where the negative sign indicates that Dimorphos's speed is reduced.} Due to the increased eccentricity, Dimorphos then begins librating and can also enter a chaotic non-principal axis (NPA) rotation state at later times depending on its shape. The attitude instability that leads to NPA rotation is driven by intersections of various spin-orbit resonances among Dimorphos's frequencies of free libration, spin precession, nutation, and mean motion --- more details can be found in \cite{Agrusa2021}. In results presented here, we give both the value for $\beta$ and the corresponding binary eccentricity, $e$, in an effort to make the results of this paper broadly applicable to other similar binary systems. Due to the non-Keplerian nature of small binary systems, we report $e$ as the geometric eccentricity, which is a function of the periapsis $(r_\text{p})$ and apoapsis $(r_\text{a})$ distances: $e =  (r_\text{a}-r_\text{p})/(r_\text{a}+r_\text{p})$.

\vspace{-5pt}
\subsection{Computation of external accelerations}
At each timestep, the \textsc{gubas} code outputs the full state of the system, including the body locations, orientations, velocities, and spins, from which the net surface accelerations of the secondary can be readily computed. The net acceleration is evaluated at the center of each triangular facet (indexed by $i$) of the shape model at each timestep (indexed by $t$) and can be written as\begin{equation}
   \mathbf{a}_{i,t}^{\text{net}} = \mathbf{a}_{i,t}^{\text{grav}} + \mathbf{a}_{i,t}^{\text{tides}} + \mathbf{a}_{i,t}^{\text{cent}} + \mathbf{a}_{i,t}^{\text{Euler}},
\end{equation}
where the vectors $\mathbf{a}_{i,t}^{\text{grav}}$,  $\mathbf{a}_{i,t}^{\text{tides}}$,  $\mathbf{a}_{i,t}^{\text{cent}}$, and $\mathbf{a}_{i,t}^{\text{Euler}}$ are the secondary's self-gravity, the primary's tidal acceleration, the centrifugal acceleration, and Euler acceleration, respectively. The Coriolis acceleration is neglected because this study is focused on the conditions to trigger surface motion, rather than details of the motion itself \citep{Kim2021}. The details of how each respective acceleration was computed can be found in Appendix \ref{app:accels}. On each facet, the surface slope is then defined as
\begin{equation}
\theta_{i,t} = \mathbf{\hat{n}}_i\cdot\mathbf{\hat{a}}_{i,t}^\text{net},
\end{equation}
where $\mathbf{\hat{n}}_i$ is the surface normal and $\mathbf{\hat{a}}_{i,t}^\text{net}=\frac{\mathbf{a}_{i,t}^{\text{net}}}{\lVert \mathbf{a}_{i,t}^{\text{net}}\rVert}$.

\vspace{-5pt}
\section{Results}
\vspace{-5pt}
\subsection{A conceptual example}
To demonstrate how the various acceleration components affect the surface slope, we show time-series plots for a scenario in which $\beta{=}2$ and $\rho_\text{S}{=}2.2\text{ g cm}^{-3}$ in Fig. \ref{fig:timeSeries}. The initial slopes of the secondary are shown in Fig.\ \ref{fig:timeSeries}(c), and the post-impact spin and orbital evolution is shown in Fig.\ \ref{fig:timeSeries}(d). The slope and accelerations are shown in Fig. \ref{fig:timeSeries}(e) for the facet shown in white in Fig.\ \ref{fig:timeSeries}(c), which has a longitude and latitude of $(\phi,\lambda) \approx (0^\circ,45^\circ$). This particular example was chosen to illustrate the relative importance of the various accelerations considered here, as well as the sensitivity of the slope evolution to the spin and orbit of Dimorphos.

The DART perturbation reduces the semimajor axis and increases the eccentricity to $e{\sim}0.023$, the effect of which can be seen in the top plot of Fig.\ \ref{fig:timeSeries}(d). Through spin-orbit coupling, Dimorphos's spin state is also excited, and it begins librating while its spin rate oscillates. In only ${\sim}5\text{ d}$, the secondary becomes attitude unstable, as indicated by the nonzero roll and pitch angles, although Dimorphos technically remains in the 1:1 spin-orbit resonance (yaw angle $<90^\circ$). The influence of these dynamical changes can be seen on the surface slope plot at the top of Fig.\ \ref{fig:timeSeries}(e). At early times, changes in the surface slope are dominated by the tidal acceleration. When Dimorphos enters slight NPA rotation, the centrifugal and Euler accelerations become much more important, leading to abrupt and chaotic surface slope changes.

\vspace{-5pt}
\subsection{Dependence on momentum enhancement (\texorpdfstring{$\beta$)})}
\begin{figure}
   \centering
   \includegraphics[clip, trim=0.225cm 0.25cm 0.25cm 0.25cm, width=0.75\hsize]{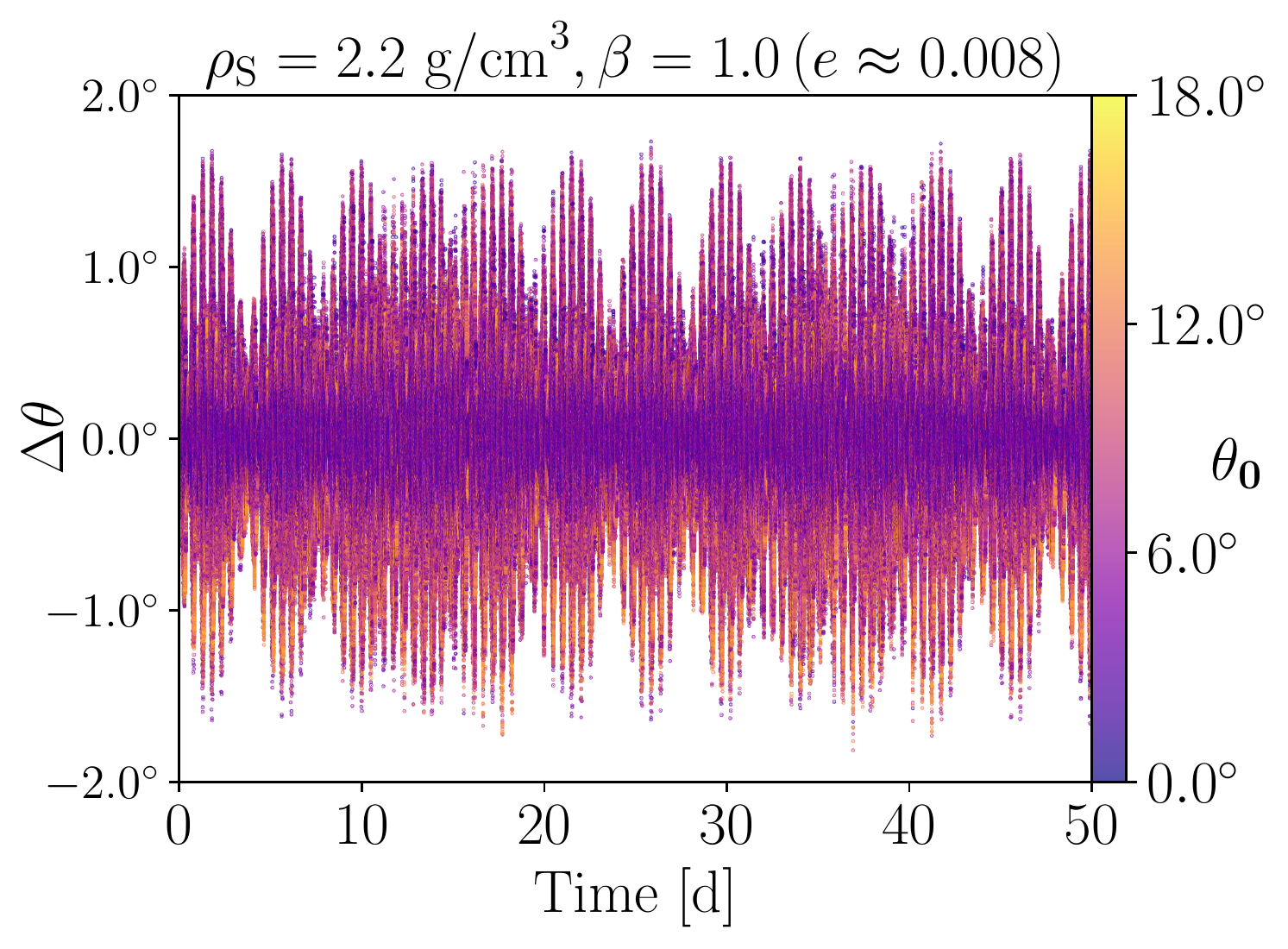}
   \includegraphics[clip, trim=0.225cm 0.25cm 0.25cm 0.25cm, width=0.75\hsize]{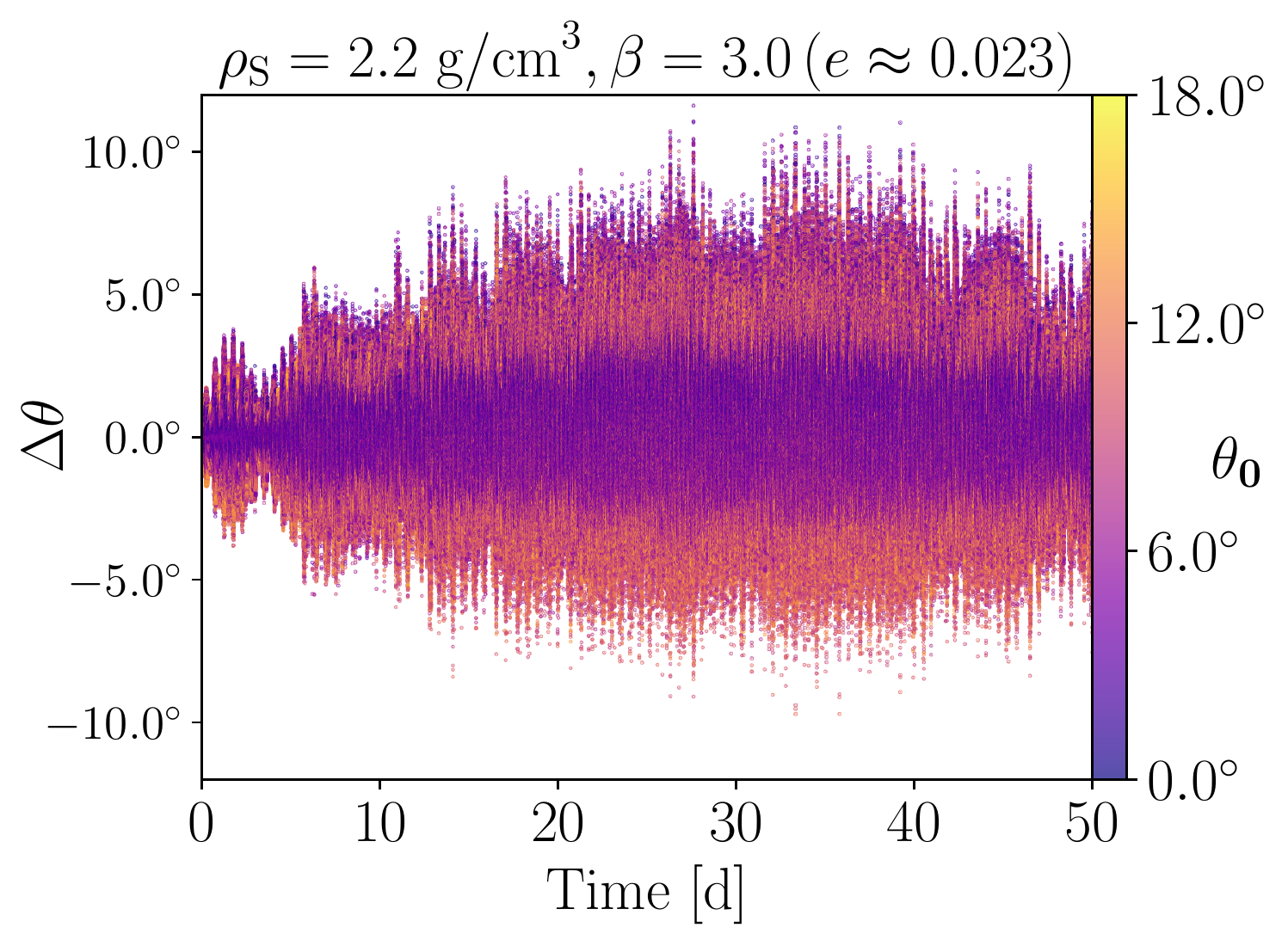}
   \caption{\label{fig:squannit_slopes_timeSeries} Time-series plots of the change in surface slope $(\Delta\theta$) of each facet in the secondary shape model. Each line is colored based on its initial surface slope $(\theta_0)$. As $\beta$ (or $e$) increases, we see much larger changes in surface slope. The bulk density is $\rho_\text{S}=2.2\text{ g cm}^{-3}$. See Appendix \ref{app:surfaceSlopePlots} for equivalent plots showing the full 365 d simulation and additional values for $\beta$.}
\end{figure}

\begin{figure}
   \centering
   \includegraphics[clip, trim=8.5cm 0.75cm 9.5cm 0.75cm,width=\hsize]{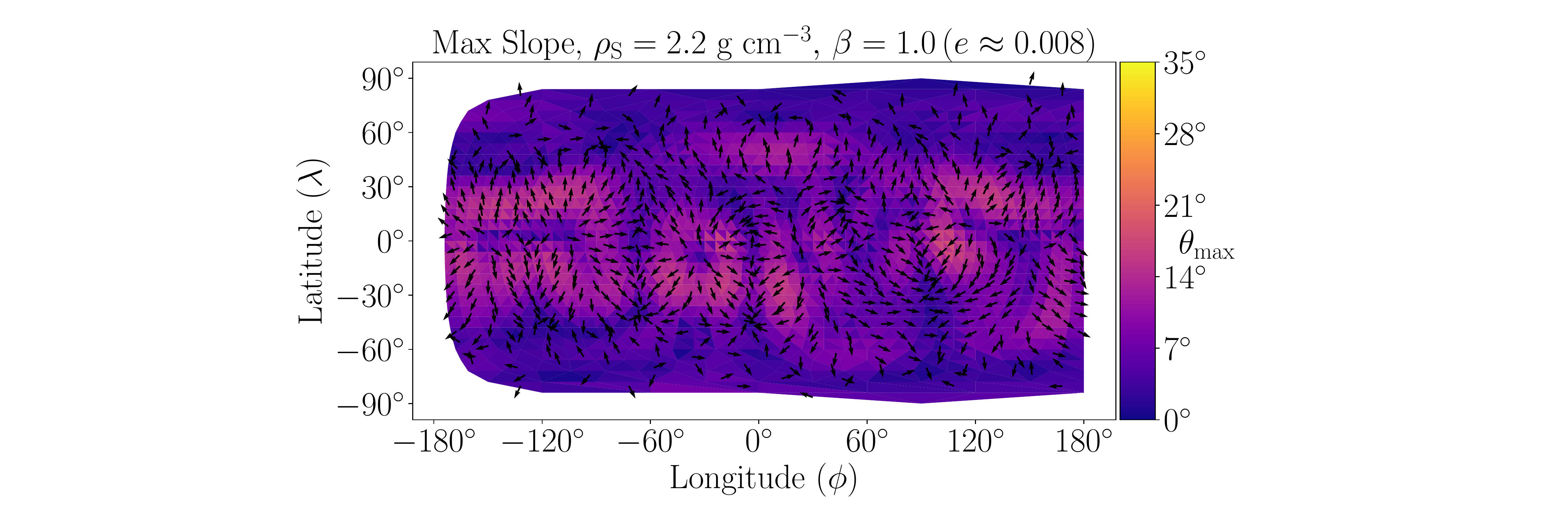}
   \includegraphics[clip, trim=8.5cm 0.75cm 9.5cm 0.75cm,width=\hsize]{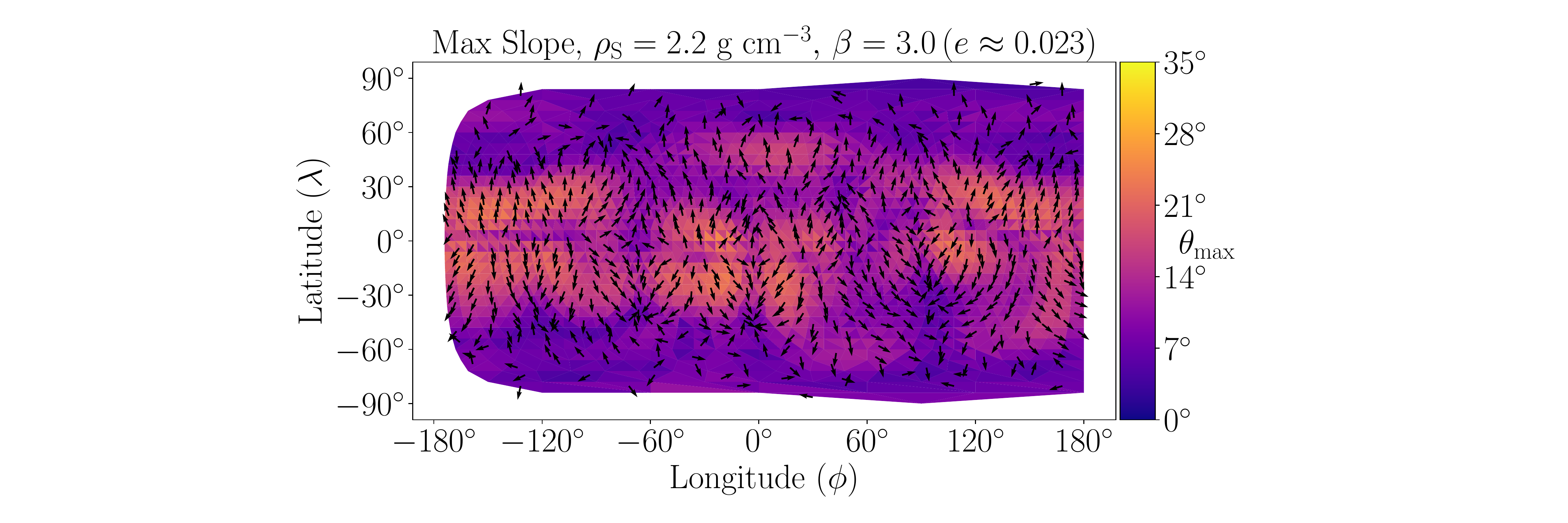}
   \caption{\label{fig:squannit_slopes_maps} Maximum slope achieved after a $365\textrm{ d}$ simulation, with arrows indicating the down-slope direction. See Appendix \ref{app:surfaceSlopePlots} for equivalent plots for other values of $\beta$.}
\end{figure}
In Fig.\ \ref{fig:squannit_slopes_timeSeries} we show time-series plots of the change in surface slope $(\Delta\theta = \theta(t) - \theta_0)$ of each surface facet for $\beta=1$ and $\beta=3$ with $\rho_\text{S}$ fixed at $2.2 \text{ g cm}^{-3}$. The color of each line corresponds to the slope at the start of the simulation, $\theta_0$. When $\beta=1$, the orbit is not significantly perturbed.\ As such, the tidal acceleration is weak and Dimorphos exhibits little NPA rotation, resulting in small surface slope changes of $\Delta\theta\lessapprox2^\circ$. When $\beta=3$, then the tidal environment becomes strong and Dimorphos enters NPA after only ${\sim}5$ d, resulting in surface slope changes as large as $\Delta\theta{\sim}10^\circ$.

The results of Fig.\ \ref{fig:squannit_slopes_timeSeries} highlight the strong temporal dependence of the surface slopes. The surface slope evolution is also spatially dependent, as demonstrated by Fig.\ \ref{fig:squannit_slopes_maps}. These plots show the maximum slope achieved over the same simulations shown in Fig.\ \ref{fig:squannit_slopes_timeSeries}. The arrows on the plot indicate the down-slope direction. These plots suggest that the highest slopes are achieved in regions that start off with a high slope. For this particular shape and assuming loose regolith covering the surface, we would expect most motion near the equator and mid-latitudes, and very little, if any, near the poles. This spatial dependence may have implications for inferred crater ages in different regions of Dimorphos's surface. In addition, the spatial dependence on the surface slope evolution could be leveraged to distinguish between causes of surface refreshment. For example, we might expect surface motion triggered by the re-accretion of impact ejecta to occur over much of Dimorphos's surface, while tidal and rotationally induced surface motion may be restricted to regions that can achieve high slopes. We refer the reader to Appendix \ref{app:surfaceSlopePlots} for additional plots that show the surface slopes for other values of $\beta$.

\vspace{-5pt}
\subsection{Dependence on the bulk density (\texorpdfstring{$\rho_\text{S}$}))}
\begin{figure}
\centering
\includegraphics[width=0.5\textwidth]{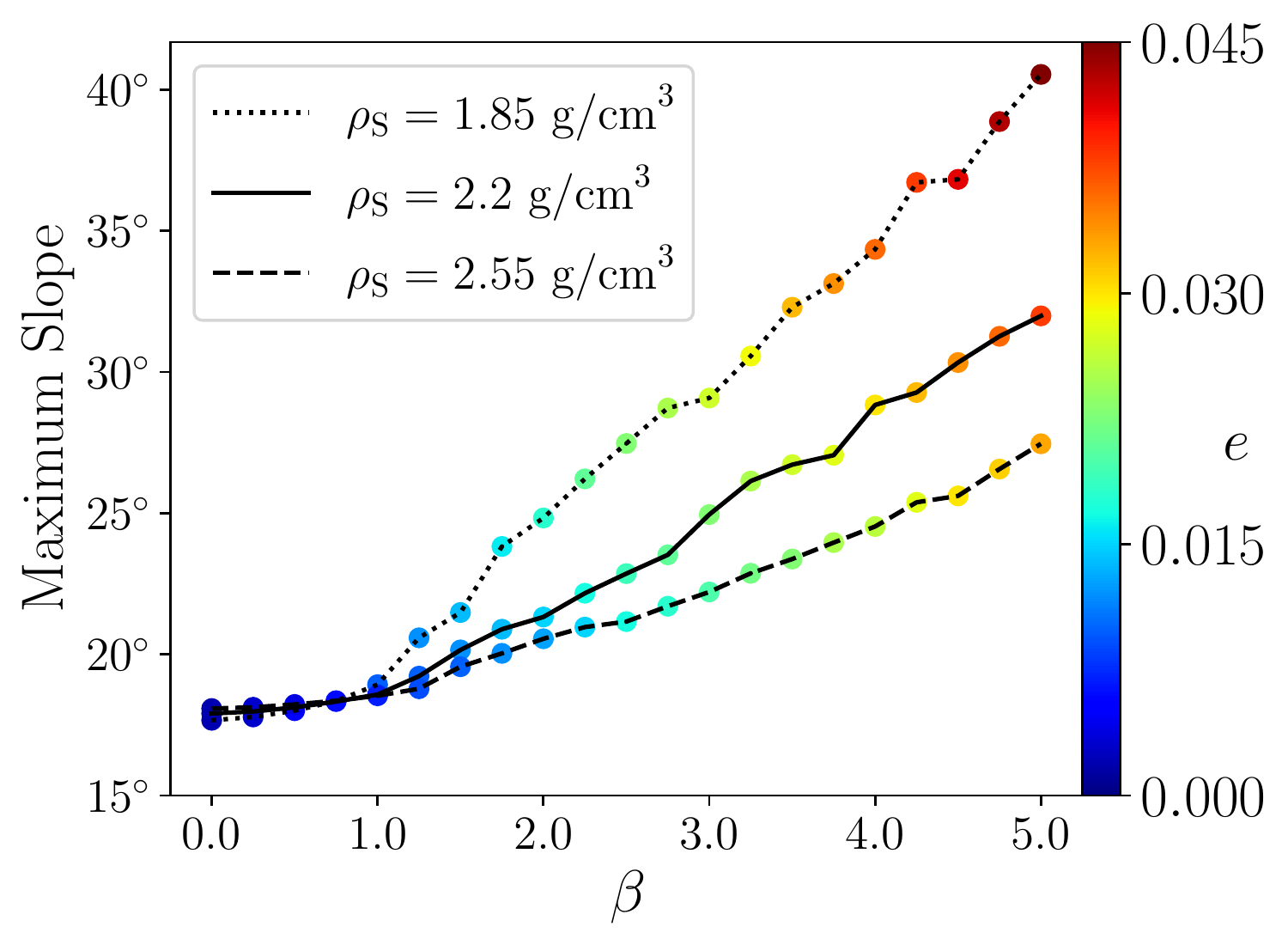}
\caption{\label{fig:maxSlope} Maximum slope as a function of $\beta$ and $\rho_\text{S}$. The slope over a given simulation increases with $\beta$ due to the spin and orbit of Dimorphos being more excited. Lower densities achieve higher slopes due to the higher orbital eccentricity for a given $\beta$, in addition to a weaker self-gravity in relation to the tidal and rotational accelerations.}
\end{figure}

The surface slopes of a given shape are highly dependent on the body's bulk density \citep{RichardsonJ2014,Susorney2022}. It sets the mass and self-gravity, which partially determine the initial slope of each facet. On a related note, a low density means that the self-gravity is weaker, making the accelerations due to tides and rotation stronger in comparison and in turn allowing larger slope changes. Finally, a low density (i.e., a low mass) means a higher eccentricity (and shorter periapsis distance) for a fixed value of $\beta$. Therefore, a lower density will result in a more perturbed orbit, in which the tidal and rotational accelerations play an increasingly important role. For these reasons, the possibility and magnitude of any granular motion will by highly dependent on Dimorphos's bulk density.

We see precisely this result in Fig.\ \ref{fig:maxSlope}, which shows the maximum surface slope achieved as a function of $\rho_\text{S}$ and $\beta$. The color of the dots indicates the eccentricity of the particular orbit, which depends on both $\beta$ and $\rho_\text{S}$. We see that the surface slopes increase dramatically as a function of $\beta$, especially for $\rho_\text{S}=1.85\text{ g cm}^{-3}$, reaching ${\sim}40^\circ$ for high $\beta$ due to the higher eccentricity and resulting in stronger tidal and rotational forces.

\vspace{-5pt}
\section{Discussion}
If Dimorphos's surface has an angle of repose of ${\sim}35^\circ$, similar to that reported at Ryugu and Bennu \citep{Watanabe2019,Barnouin2022}, then we would expect significant landslides and shape changes in cases where $\theta$ exceeds this value. For the Dimorphos shape used in this study, this would only occur for lower densities and high $\beta$ values. Without knowing the true shape of Dimorphos, however, it is impossible to say with certainty how probable any post-impact surface motion is. The aim of this paper is to demonstrate the plausibility of any dynamics-induced granular motion or shape change, and this topic will be revisited once Dimorphos's true shape is known.

Recent work focused on surface refreshment on Mars's moon Phobos indicates that a time-varying $\Delta\theta$ of only a few degrees can lead to a gradual creep motion of granular material, without the slope ever exceeding the formal angle of repose. \cite{Ballouz2019} combined dynamical modeling, granular physics, and geologic mapping of color units to demonstrate that regions of combined high values of $\theta$ and $\Delta\theta$ coincide with Phobos's blue units. This work indicated an active surface-refreshing process that could excavate pristine un-weathered material. Depending on Dimorphos's geophysical properties, it may be plausible that a similar creep motion process will occur following the DART impact. We note that surface refreshment could be currently ongoing, if Dimorphos is already in an NPA rotation state as predicted by \cite{Quillen2022a}.

It is also important to consider that both $\beta$ and $\rho_\text{S}$ could lie outside the range explored in this paper. Of course, Dimorphos's real shape and surface geology are also unknown, so the results presented here are illustrative and meant to highlight the range of post-impact possibilities. After DART's impact, this phenomenon can be explored with higher fidelity, incorporating the initial shape model and surface geology obtained with DART and LICIACube imagery. When Hera arrives, its optical instruments and CubeSats, especially the Juventas CubeSat and its onboard GRAvimeter for small Solar System bodies (GRASS) instrument, will measure the dynamical slopes as one of its science objectives \citep{Michel2018,Karatekin2021,Ritter2021}. The seismic pulse delivered by the DART impact may significantly alter surface features on Dimorphos \citep{Quillen2022b,Thomas2005}. We also note that the global shape of Dimorphos may also be immediately altered by the DART impact itself \citep{Raducan2022b}. In addition to affecting the system dynamics \citep{Nakano2022}, these processes will create a unique challenge in discerning the various surface refreshment mechanisms upon Hera's arrival.

The results of the work presented here have the following implications, in the context of the DART and Hera missions as well as binary asteroids in general:

\textbf{Granular motion and surface changes.} Through images and infrared measurements, Hera may identify refreshed areas of Dimorphos's surface exposed by dynamics-induced surface motion. Furthermore, a comparison of images taken by DART and Hera may be used to identify surface features that have moved or changed during the four years between the missions. If there is long-term boulder motion on the surface, Hera may detect the motion of boulders over the course of its six-month mission lifetime. Furthermore, this effect may noticeably alter the system's dynamics \citep{Brack2019}. 

\textbf{Crater degradation.} Impact craters (both natural craters and DART's crater) may degrade at different rates based on their location on the surface as surface slope changes are spatially dependent. This may have important implications for understanding crater morphology and the surface age of Dimorphos, a challenge that does not usually require consideration for single asteroids due to their quasi-static spin states \citep{Sugita2019,Walsh2019,RichardsonJ2020}.

\textbf{Tidal dissipation.} Granular surface motion may affect tidal dissipation in two ways. First, any material undergoing surface motion will dissipate energy through friction, potentially enhancing dissipation beyond what is assumed from traditional tidal theories \citep{Goldreich2009,Nimmo2019}. Second, granular motion will change Dimorphos's mass distribution and, therefore, its gravitational potential. This mechanism could subtly remove energy from the system, an effect not captured by simplified tidal treatments.

\textbf{Binary formation and evolution.} One proposed scenario of binary formation assumes the secondary forms through a spin-up fission event driven by the Yarkovsky–O'Keefe–Radzievskii–Paddack (YORP) effect and initially orbits chaotically. At some later time, the secondary must fission a second time, forming a short-lived triple system and liberating excess free energy in order to enter a stable, synchronous spin state \citep{Jacobson2011a}. Given the results presented herein, we might expect landslides on the surface well before a secondary fission event. This process may dissipate energy and reshape the secondary, allowing for synchronous rotation without the need to invoke additional fissions. Furthermore, if all secondaries undergo chaotic rotation at some point, then we might expect the population to have broadly similar shapes. However, this would largely depend on the relative timescales for tidal locking and surface refreshment, as well as other competing slope-altering processes such as meteorite impacts. In any case, rotation-driven surface motion, shape change, and energy dissipation may be important effects that should be accounted for in any binary asteroid formation scenario.

\vspace{-5pt}
\section{Conclusions}
In this paper we have shown that perturbed post-impact spin and orbital dynamics may lead to significant fluctuations in Dimorphos's surface slopes. Depending on Dimorphos's shape, bulk density, surface geology, and $\beta$, we predict that this may trigger long-lived granular motion on the surface. The implications for dynamics-driven granular motion include a refreshment of Dimorphos's surface, impact crater degradation, and enhanced tidal dissipation. Understanding these effects will help guide and interpret the measurements Hera will obtain on Dimorphos's surface and interior. In addition, this effect may have implications for the formation and evolution of small binary systems in general.

Thanks to this initial study, post-impact granular motion will be explored more closely and with higher fidelity when Dimorphos's shape model first becomes available. Future work includes directly modeling granular motion on the surface in addition to coupling that motion back to the resulting dynamical evolution.

\begin{acknowledgements}
   We kindly thank the anonymous referee, whose insightful comments significantly improved the manuscript.
   This work was supported in part by the DART mission, NASA Contract \#80MSFC20D0004 to JHU/APL. E.T., G.N., O.K., and P.M. acknowledge support from the European Union’s Horizon 2020 research and innovation program under grant agreement No. 870377 (project NEO-MAPP). E.T., G.N., and O.K. acknowledge support by Belgian Federal Science Policy (BELSPO) through the through the ESA/PRODEX Program. P.M. acknowledges support from ESA and from the French Space Agency CNES. The simulations herein were carried out on The University of Maryland Astronomy Department’s YORP cluster, administered by the Center for Theory and Computation.
\end{acknowledgements}

%
\bibliographystyle{aa} 
\bibliography{references} 
%

\begin{appendix} 
\vspace{-5pt}
\section{Calculation of surface accelerations and slopes}\label{app:accels}

We provide additional details for how exactly the accelerations were computed over the surface of the secondary shape model. On each facet, all accelerations were evaluated at the midpoint (i.e., the center) of the given facet. 

\vspace{-5pt}
\subsection{Gravitational accelerations}
On a given facet, $i$, and at a given time, $t$, the two gravitational accelerations felt on the surface are due to self-gravity, $\mathbf{a}_{i,t}^\text{grav}$, and the tidal acceleration due to the presence of the primary, $\mathbf{a}_{i,t}^\text{tides}$. 

The self-gravity was computed using an algorithm identical to that presented in \cite{Werner1997}. This method computes the {exact} gravitational acceleration due to a polyhedral shape model with uniform density. Though the calculation is somewhat computationally expensive, it only needs to be done once, as we assume that Dimorphos's global shape does not undergo significant change. At each facet, $\mathbf{a}_{i,t}^\text{grav}$ was computed in Dimorphos's body-fixed frame.

Unlike the self-gravity, the tidal acceleration must be computed at every facet at every timestep. Therefore, we turned to MacCullagh's formula to approximate the tides to save computational costs while still capturing effects due to Didymos's irregular shape. MacCullagh's formula is written as \citep{MacCullagh1844a, MacCullagh1844b, Murray2000}

\begin{equation}
   V = \frac{GM}{r} - \frac{GM}{2r^5}f(A,B,C,x,y,z),
\end{equation}
where $G$ is the gravitational constant, $M$ is the body mass, $r$ is the distance from the body's barycenter to the external field point, $A$, $B$, and $C$ are the body's principal moments of inertia, $x$, $y$, and $z$ are the coordinates of the external field point measured in the primary body-fixed frame, and $f$ is defined as\begin{equation}
   f(A,B,C,x,y,z) = (B+C-2A)x^2 + (C+A-2B)y^2 + (A+B-2C)z^2.
\end{equation}

The gravitational acceleration can be calculated by taking partial derivatives of $V$ with respect to $x$, $y$, and $z$ (see Chapter 5 of \cite{Murray2000}):

\begin{align}
a_{x} &= -\frac{\partial V}{\partial x} = -\frac{GMx}{r^3}+\frac{G(B+C-2A)x}{r^5} - \frac{5Gx}{2r^7}f(A,B,C,x,y,z)\\
a_{y} &= -\frac{\partial V}{\partial y} = -\frac{GMy}{r^3}+\frac{G(A+C-2B)y}{r^5} - \frac{5Gy}{2r^7}f(A,B,C,x,y,z)\\
a_{z} &= -\frac{\partial V}{\partial z} = -\frac{GMz}{r^3}+\frac{G(A+B-2C)z}{r^5} - \frac{5Gz}{2r^7}f(A,B,C,x,y,z).
\end{align}

In order to calculate the net gravitational acceleration felt at a point on Dimorphos's surface due to Didymos, we took the difference between the acceleration evaluated at a given surface point and the acceleration evaluated at Dimorphos's barycenter. This acceleration vector was computed in the primary's body-fixed frame before being rotated into the secondary's body-fixed frame.

\vspace{-5pt}
\subsection{Rotational accelerations}
The centrifugal acceleration at a given point on Dimorphos's surface denoted with the index $i$ can be written as
\begin{equation}
   \mathbf{a}_{i,t}^{cent} = (\mathbf{\Omega}_t\times\mathbf{r}_{i})\times\mathbf{\Omega}_t,
\end{equation}
where $\mathbf{\Omega}_t$ is the spin angular velocity vector of Dimorphos at a given time and $\mathbf{r_i}$ is the position vector of the surface point, coordinated in Dimorphos's body-fixed frame. Since Dimorphos's spin rate is time varying, we also account for the Euler acceleration,
\begin{equation}
   \mathbf{a}_{i,t}^{\text{Euler}} = \mathbf{r}_{i}\times\frac{d\mathbf{\Omega}_t}{dt}.
 \end{equation} 

Since \textsc{gubas} does not directly output the time derivative of the secondary's spin, we calculated it in post-processing with a fourth-order central finite-difference scheme. At a 60 second timestep, the fourth-order approximation sufficiently approximates $\frac{d\mathbf{\Omega}}{dt}$.

\vspace{-5pt}
\section{Longer-term spin-orbit and surface slope evolution for the test case \label{app:timeSeries}}
\begin{figure*}
      \centering    
      \begin{minipage}[b]{0.45\hsize}
         \centering
         \includegraphics[width=\textwidth]{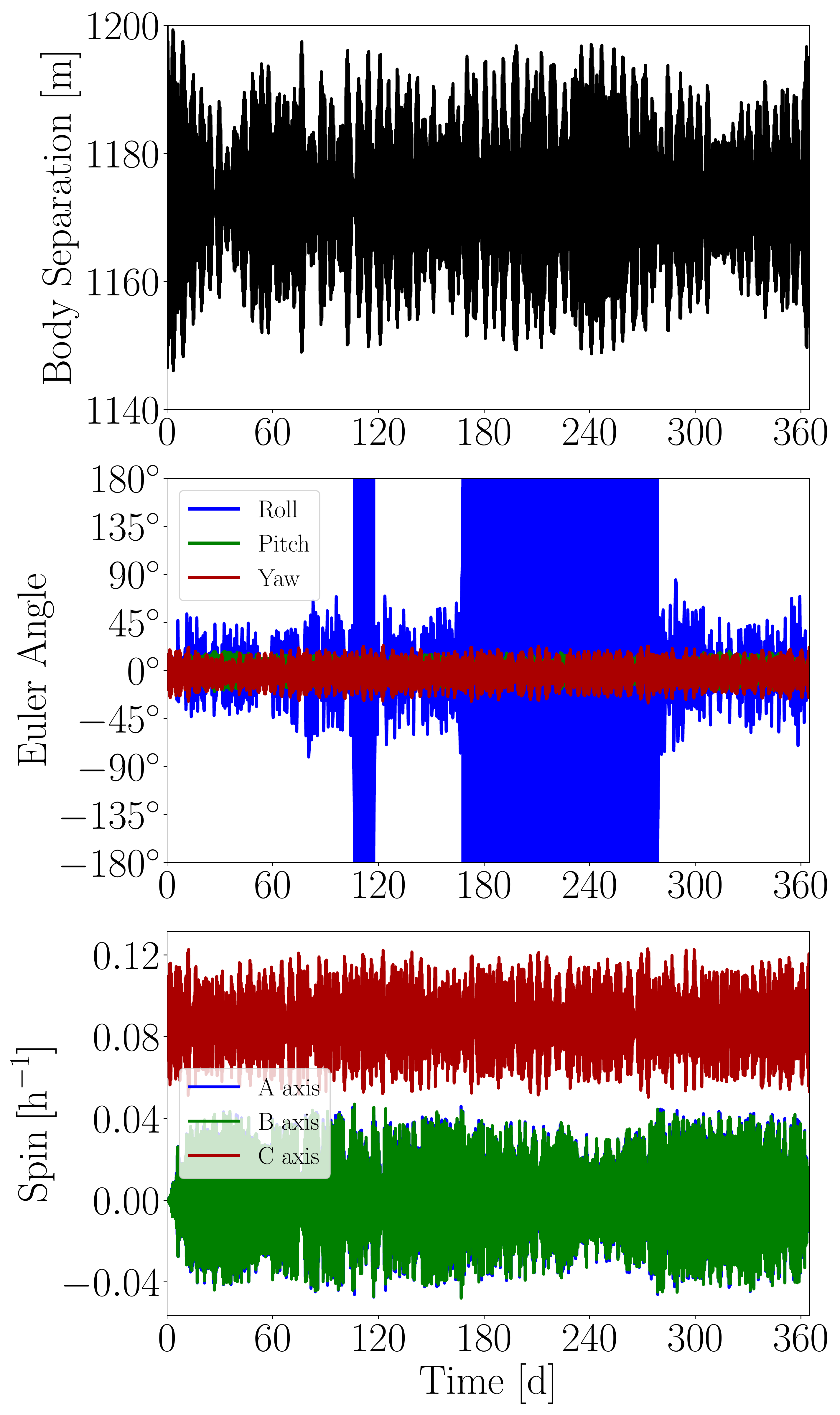}
         (a)
      \end{minipage}
      \begin{minipage}[b]{0.45\hsize}
         \centering
         \includegraphics[width=\textwidth]{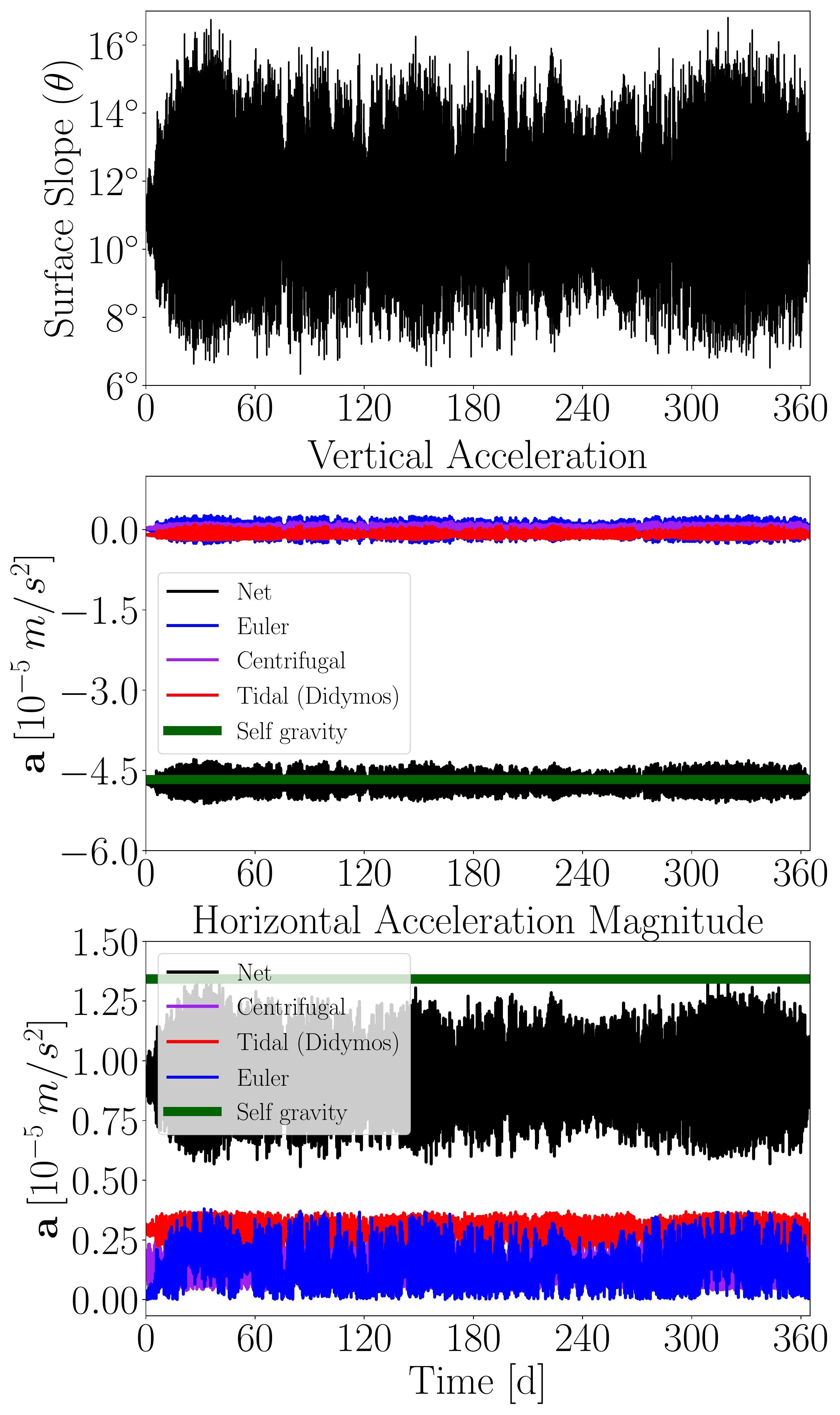}
         (b)
      \end{minipage}
      \caption{\label{app:fig:timeSeries}
      Spin, orbit, and surface slope evolution for the Squannit-shaped Dimorphos when $\beta=3\, (e=0.023)$. (a) Body separation, Euler angles, and body spin of Dimorphos. (b) Slope and surface accelerations on a facet near  $(\phi, \lambda)\approx(0^\circ,45^\circ)$. These plots are identical to those of Fig.\ 1, coming from the same simulation, except they show a longer time duration to highlight how the evolving spin and orbital motion of Dimorphos influences the accelerations felt on the surface. Dimorphos's NPA rotation (as indicated by the roll and pitch angles or spin about the A and B axes) leads to large increases in the centrifugal and Euler accelerations that are capable of driving large surface slope changes.}
   \end{figure*}

Figure \ref{app:fig:timeSeries} shows plots identical to those in Fig.\ \ref{fig:timeSeries} but with the time span increased to 365 d in order to show the longer-term evolution of the dynamics and slope. These plots indicate that the dominant mechanism for changing the surface slope is the NPA rotation of the secondary. We also see that Dimorphos is able to enter the ``barrel instability,'' a unique spin state where the secondary remains tidally locked despite rolling about its long axis, as indicated by the roll angle hitting $180^\circ$ \citep{Cuk2021}. It seems that this does not significantly affect the surface slopes, however.

\vspace{-5pt}
\section{Additional surface slope plots \label{app:surfaceSlopePlots}}
\begin{figure*}
      \centering
      \begin{minipage}[b]{0.4\hsize}
         \centering
         \includegraphics[width=\textwidth]{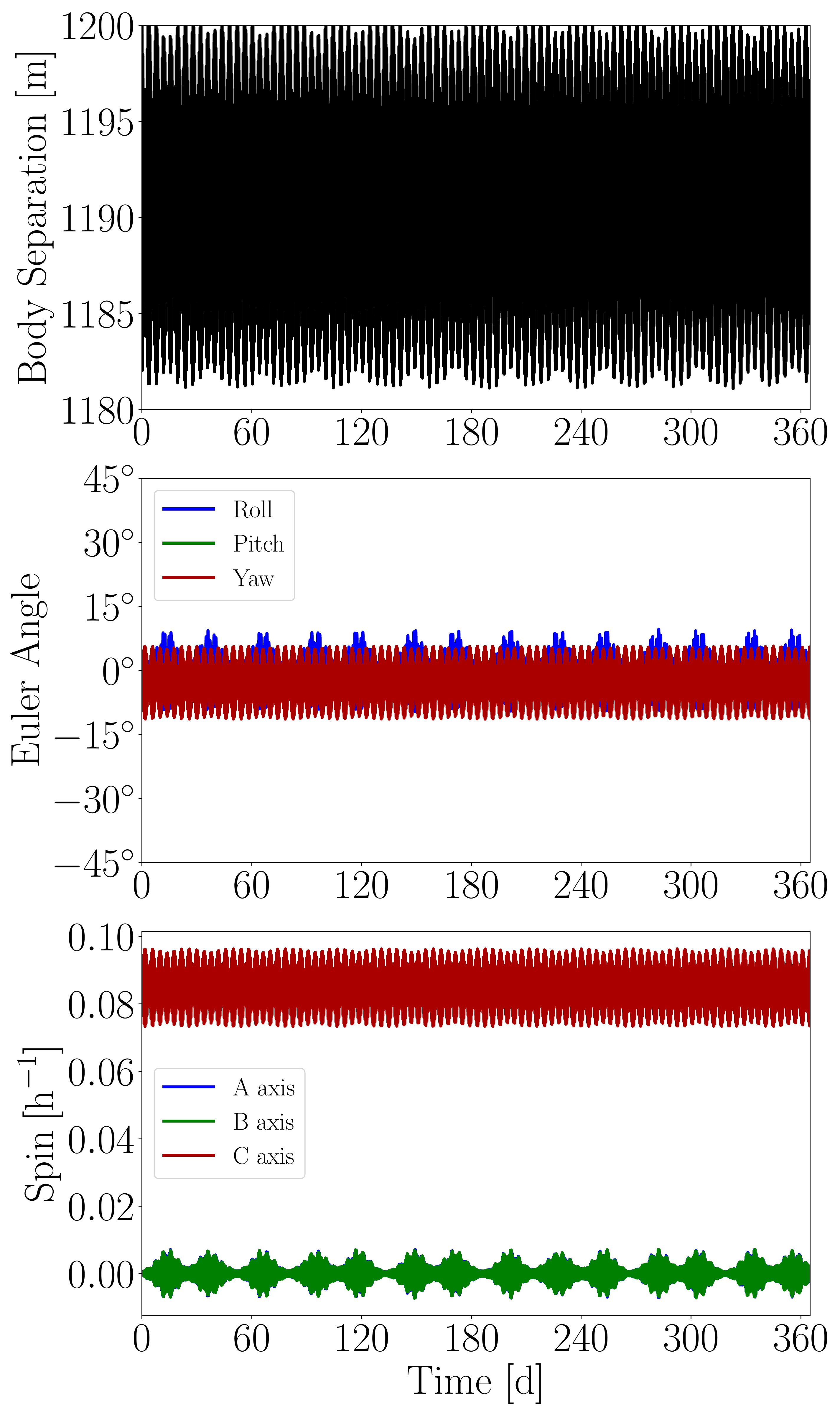}
         (a)
      \end{minipage}
      \begin{minipage}[b]{0.55\hsize}
         \centering
         \includegraphics[clip, trim=0.225cm 0.25cm 0.25cm 0.25cm,width=\textwidth]{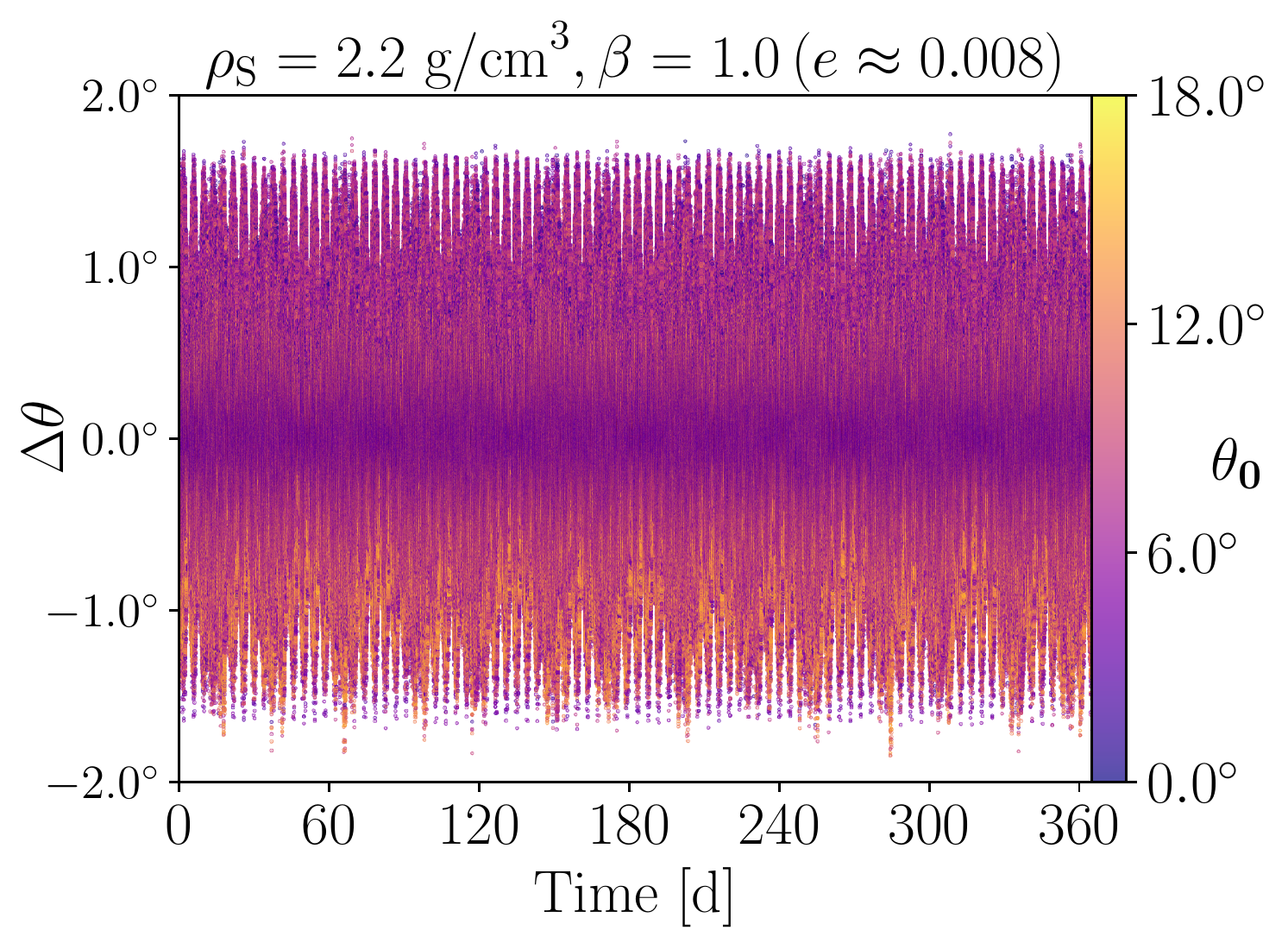}
         (b) 
         \includegraphics[clip, trim=8.5cm 0.75cm 9.5cm 0.75cm, width=\textwidth]{slopeMap_maxSlope_rho2.2_beta1.pdf}
         (c)
      \end{minipage}        
      \caption{\label{app:fig:beta1} The spin, orbit, and surface slope evolution for $\beta=1$ and $\rho_\text{S}=2.2 \textrm{ g cm}^{-3}$ over 365 days. (a) Body separation, Euler angles, and body spin of Dimorphos. (b) Corresponding change in slope ($\Delta\theta$) over time for each surface facet, colored by the starting slope ($\theta_0$) of that facet. (c) Maximum surface slope achieved on each facet over the full 365 d simulation, with arrows pointing in the down-slope direction. }
   \end{figure*}

\begin{figure*}
      \centering
    
      \begin{minipage}[b]{0.4\hsize}
         \centering
         \includegraphics[width=\textwidth]{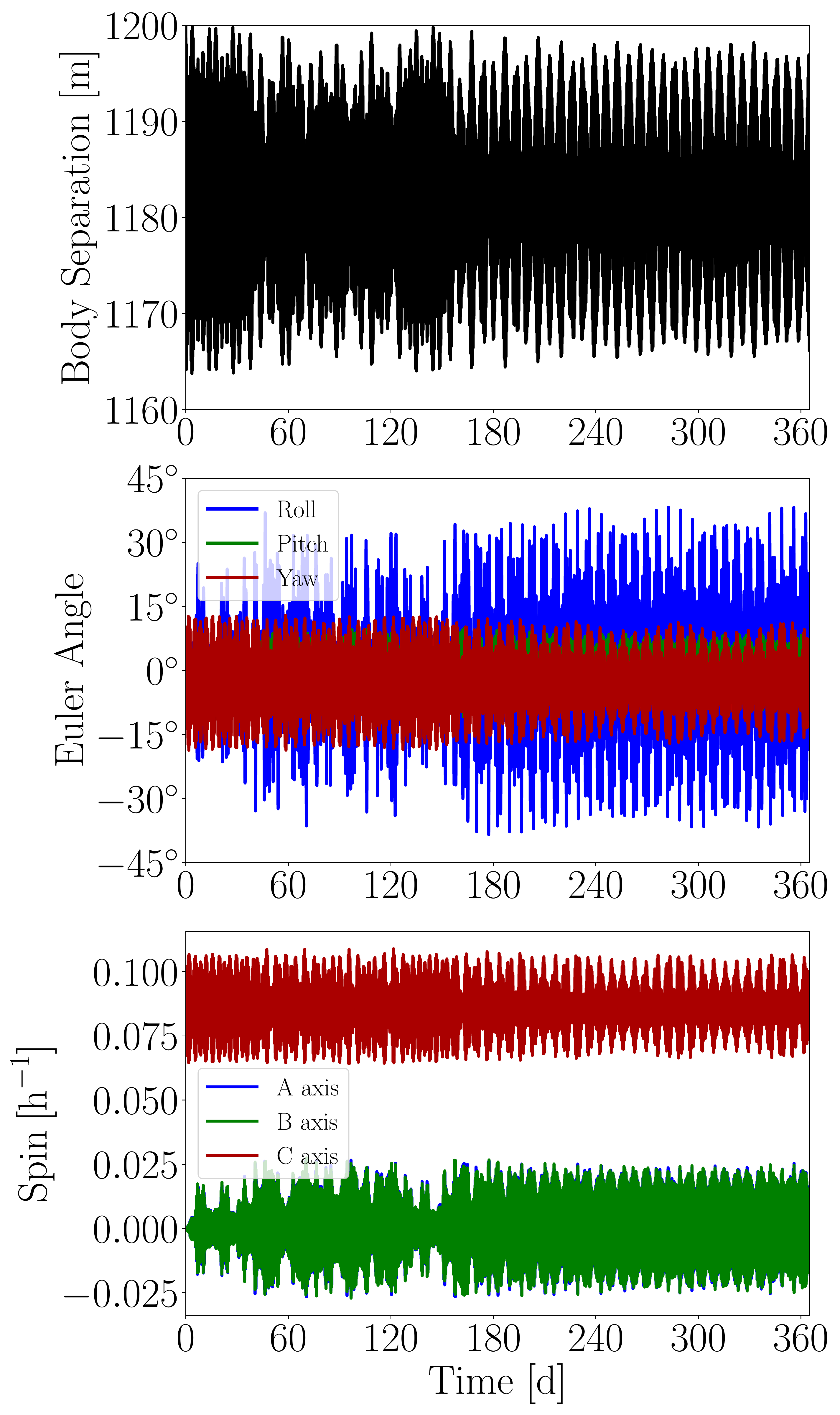}
         (a)
      \end{minipage}
      \begin{minipage}[b]{0.55\hsize}
         \centering
         \includegraphics[clip, trim=0.225cm 0.25cm 0.25cm 0.25cm,width=\textwidth]{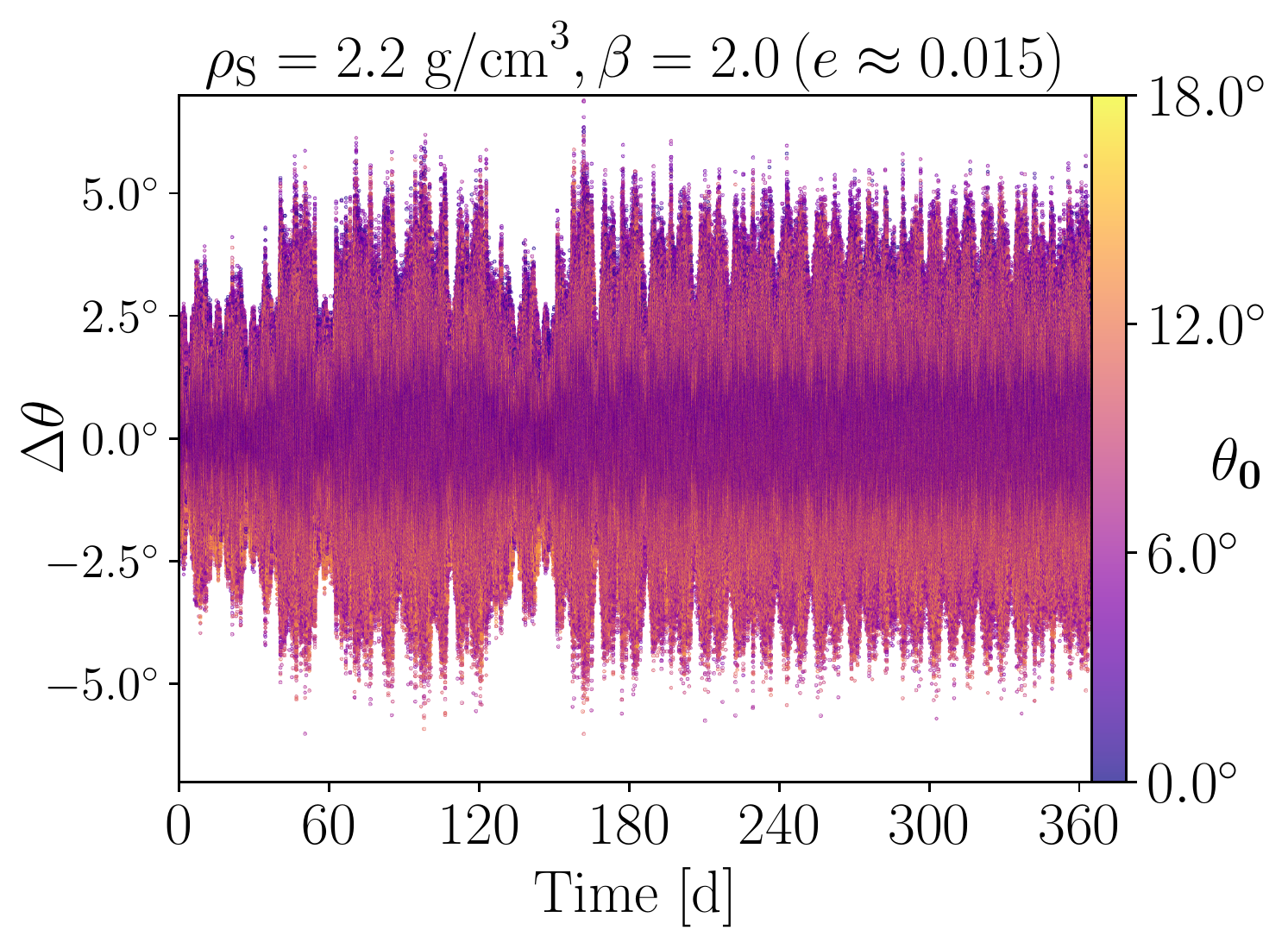}
         (b) 
         \includegraphics[clip, trim=8.5cm 0.75cm 9.5cm 0.75cm, width=\textwidth]{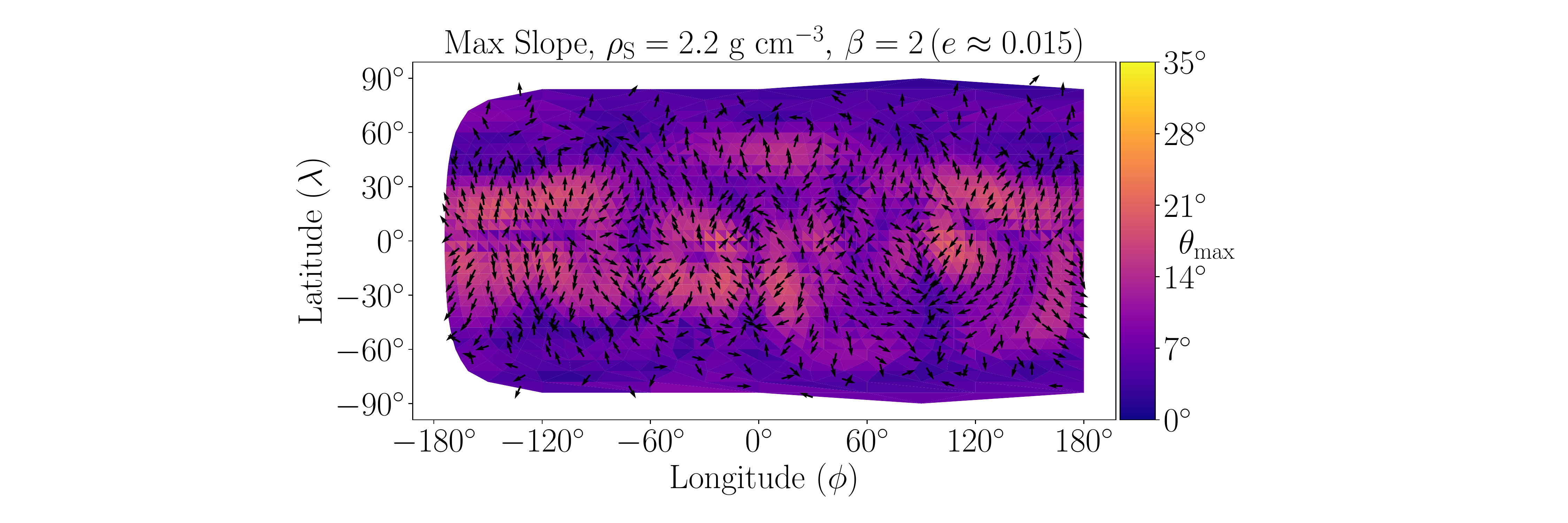}
         (c)
      \end{minipage}        
      \caption{\label{app:fig:beta2}  The spin, orbit, and surface slope evolution for $\beta=2$ and $\rho_\text{S}=2.2 \textrm{ g cm}^{-3}$ over 365 days. (a) Body separation, Euler angles, and body spin of Dimorphos. (b) Corresponding change in slope ($\Delta\theta$) over time for each surface facet, colored by the starting slope ($\theta_0$) of that facet. The spikes in $\Delta\theta$ correspond to periods of increased NPA rotation of Dimorphos. (c) Maximum surface slope achieved on each facet over the full 365 d simulation, with arrows pointing in the down-slope direction. }
   \end{figure*}

\begin{figure*}
      \centering
    
      \begin{minipage}[b]{0.4\hsize}
         \centering
         \includegraphics[width=\textwidth]{spinOrbit_1yr_rho2.2_beta3.pdf}
         (a)
      \end{minipage}
      \begin{minipage}[b]{0.55\hsize}
         \centering
         \includegraphics[clip, trim=0.225cm 0.25cm 0.25cm 0.25cm,width=\textwidth]{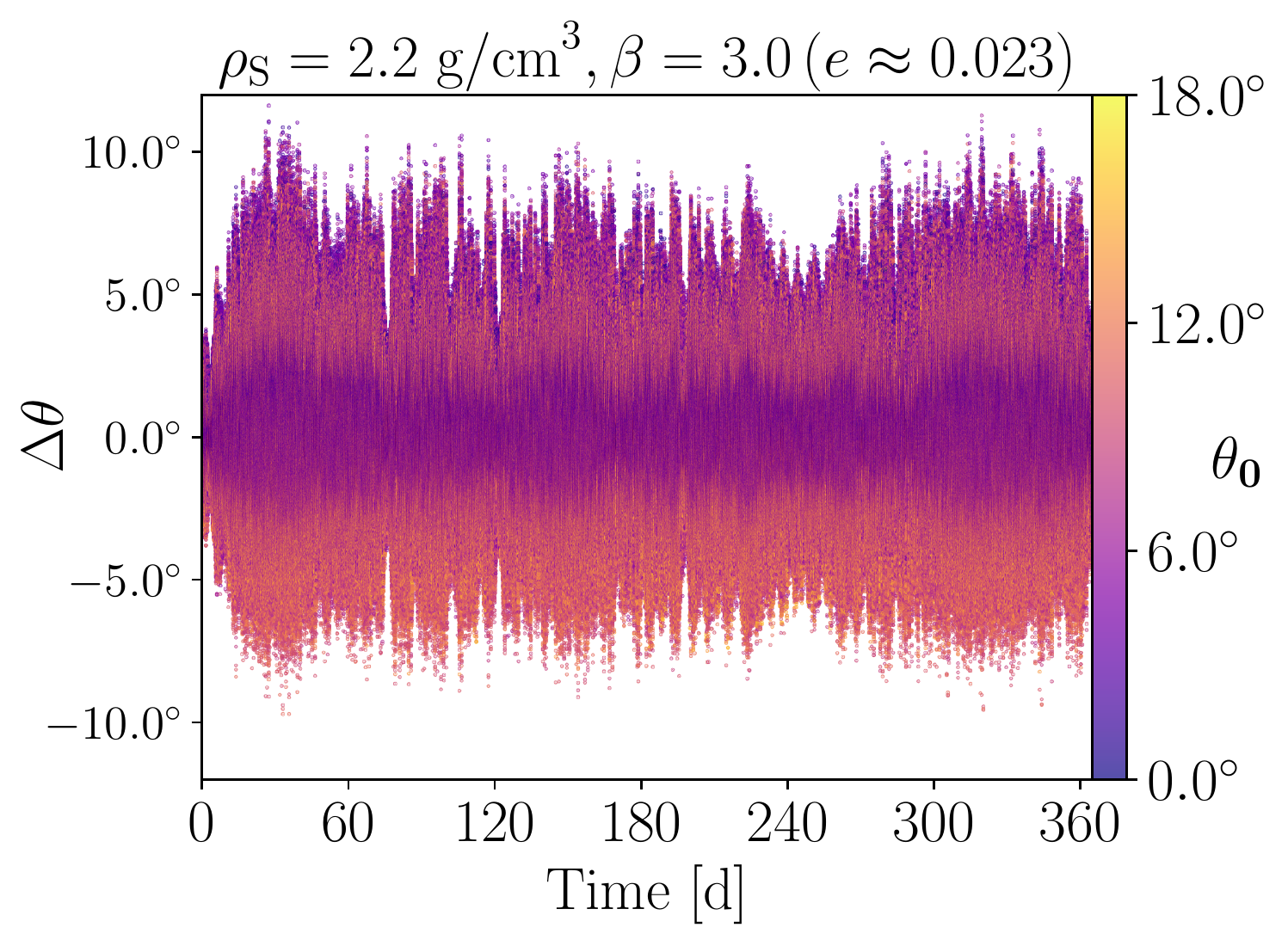}
         (b) 
         \includegraphics[clip, trim=8.5cm 0.75cm 9.5cm 0.75cm, width=\textwidth]{slopeMap_maxSlope_rho2.2_beta3.pdf}
         (c)
      \end{minipage}        
      \caption{\label{app:fig:beta3} The spin, orbit, and surface slope evolution for $\beta=3$ and $\rho_\text{S}=2.2 \textrm{ g cm}^{-3}$ over 365 days. (a) Body separation, Euler angles, and body spin of Dimorphos. (b) Corresponding change in slope ($\Delta\theta$) over time for each surface facet, colored by the starting slope ($\theta_0$) of that facet. The surface slope evolution is dominated by NPA rotation. (c) Maximum surface slope achieved on each facet over the full 365 d simulation, with arrows pointing in the down-slope direction. }
   \end{figure*}

\begin{figure*}
      \centering
    
      \begin{minipage}[b]{0.4\hsize}
         \centering
         \includegraphics[width=\textwidth]{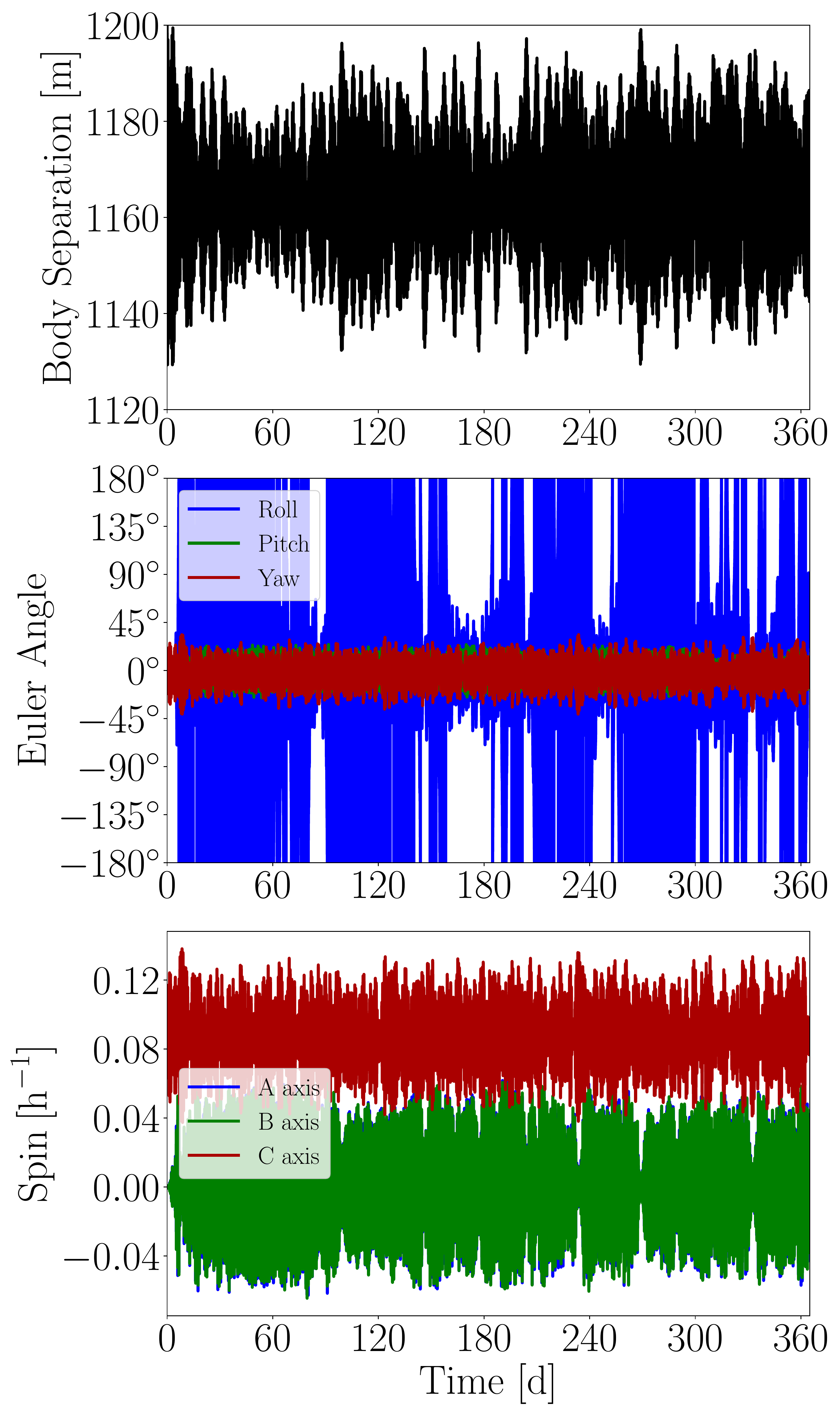}
         (a)
      \end{minipage}
      \begin{minipage}[b]{0.55\hsize}
         \centering
         \includegraphics[clip, trim=0.225cm 0.25cm 0.25cm 0.25cm,width=\textwidth]{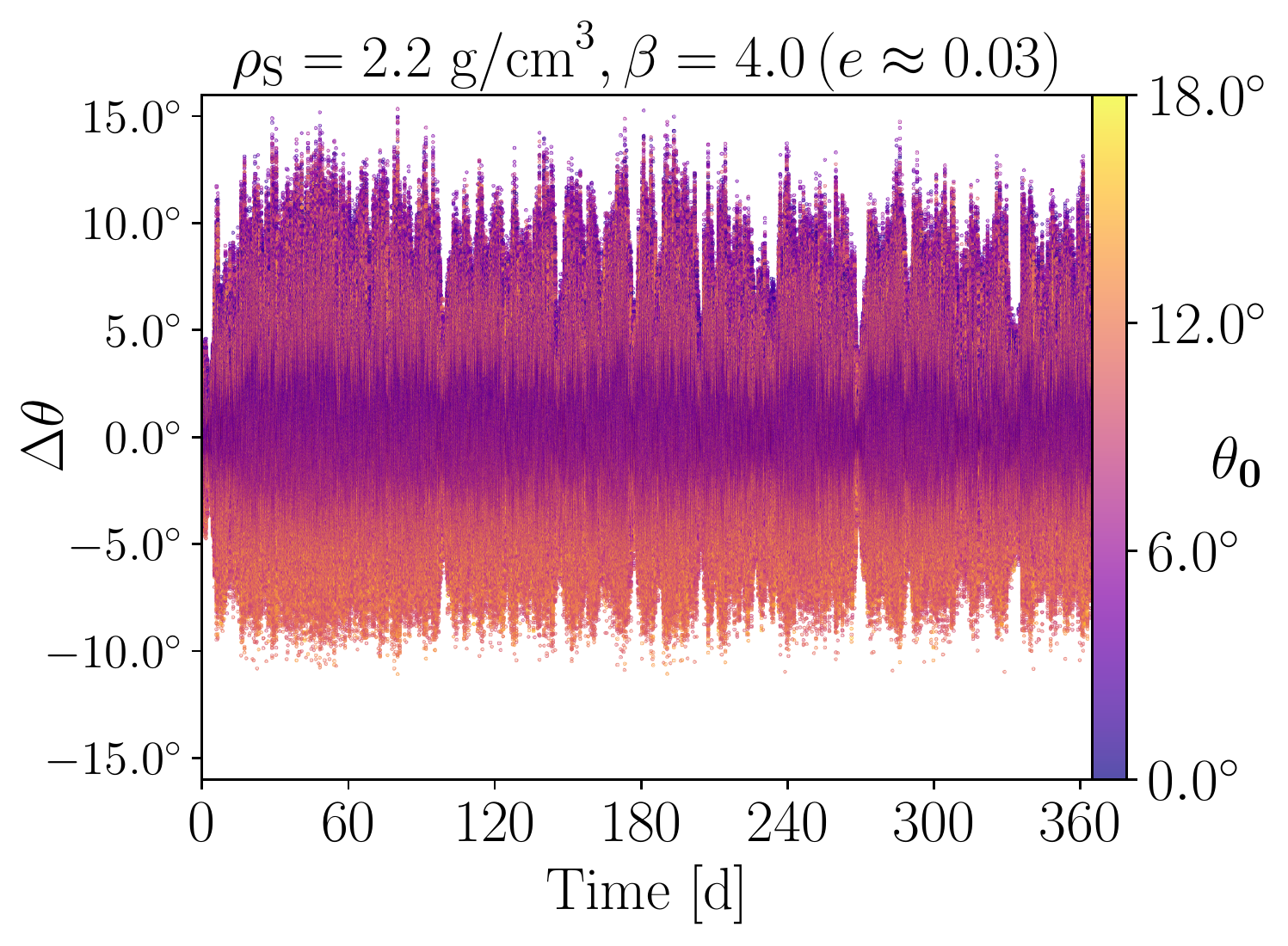}
         (b) 
         \includegraphics[clip, trim=8.5cm 0.75cm 9.5cm 0.75cm, width=\textwidth]{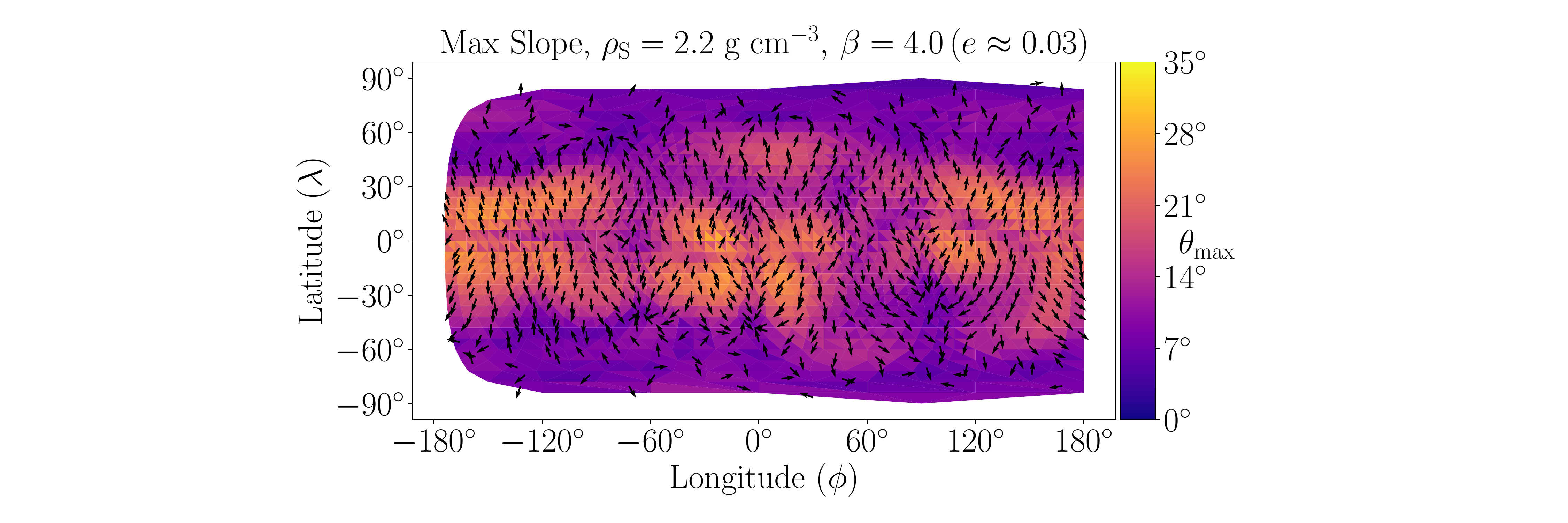}
         (c)
      \end{minipage}        
      \caption{\label{app:fig:beta4}  The spin, orbit, and surface slope evolution for $\beta=4$ and $\rho_\text{S}=2.2 \textrm{ g cm}^{-3}$ over 365 days. (a) Body separation, Euler angles, and body spin of Dimorphos. (b) Corresponding change in slope ($\Delta\theta$) over time for each surface facet, colored by the starting slope ($\theta_0$) of that facet. The surface slope evolution is dominated by NPA rotation. (c) Maximum surface slope achieved on each facet over the full 365 d simulation, with arrows pointing in the down-slope direction. }
   \end{figure*}

\begin{figure*}
      \centering
    
      \begin{minipage}[b]{0.4\hsize}
         \centering
         \includegraphics[width=\textwidth]{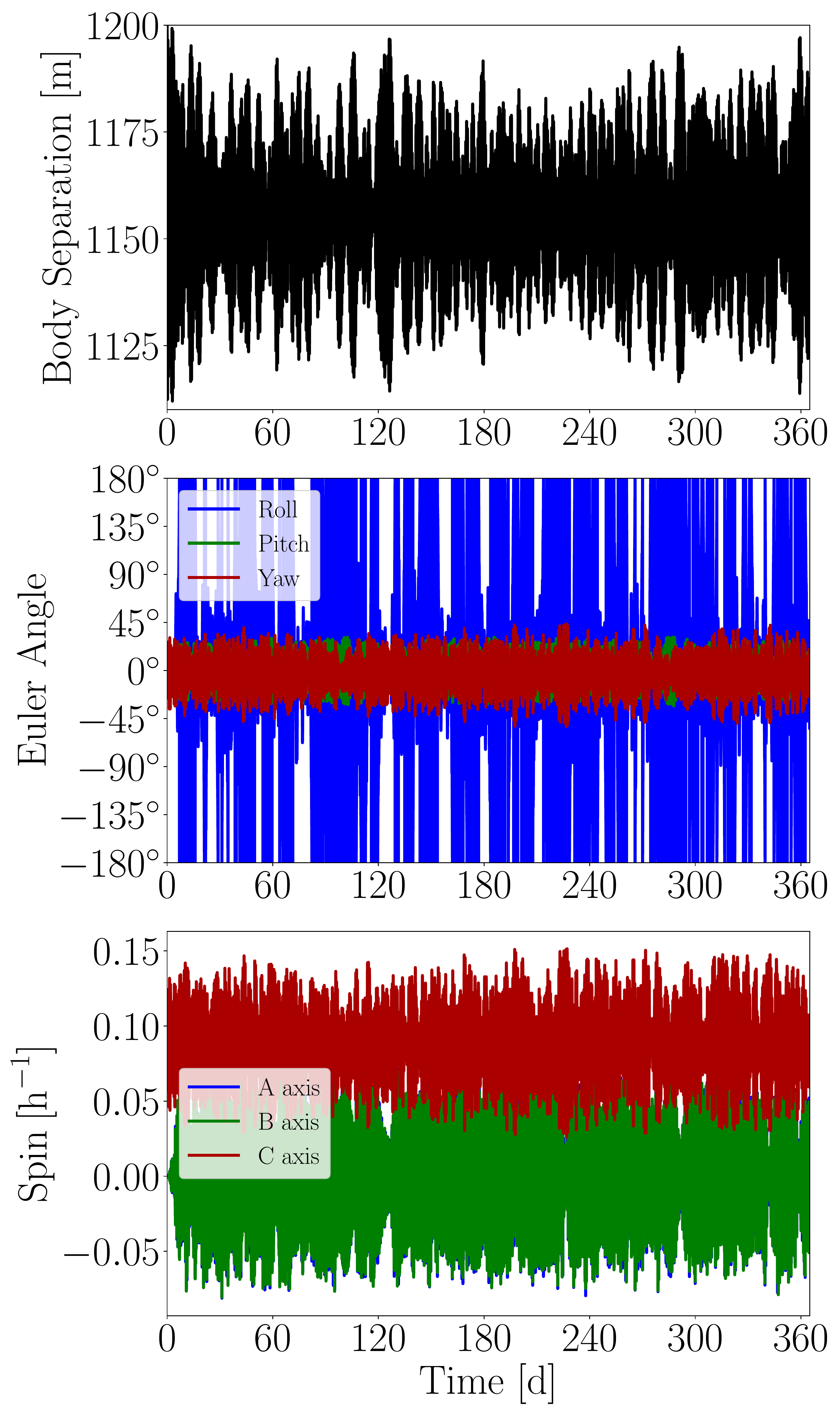}
         (a)
      \end{minipage}
      \begin{minipage}[b]{0.55\hsize}
         \centering
         \includegraphics[clip, trim=0.225cm 0.25cm 0.25cm 0.25cm,width=\textwidth]{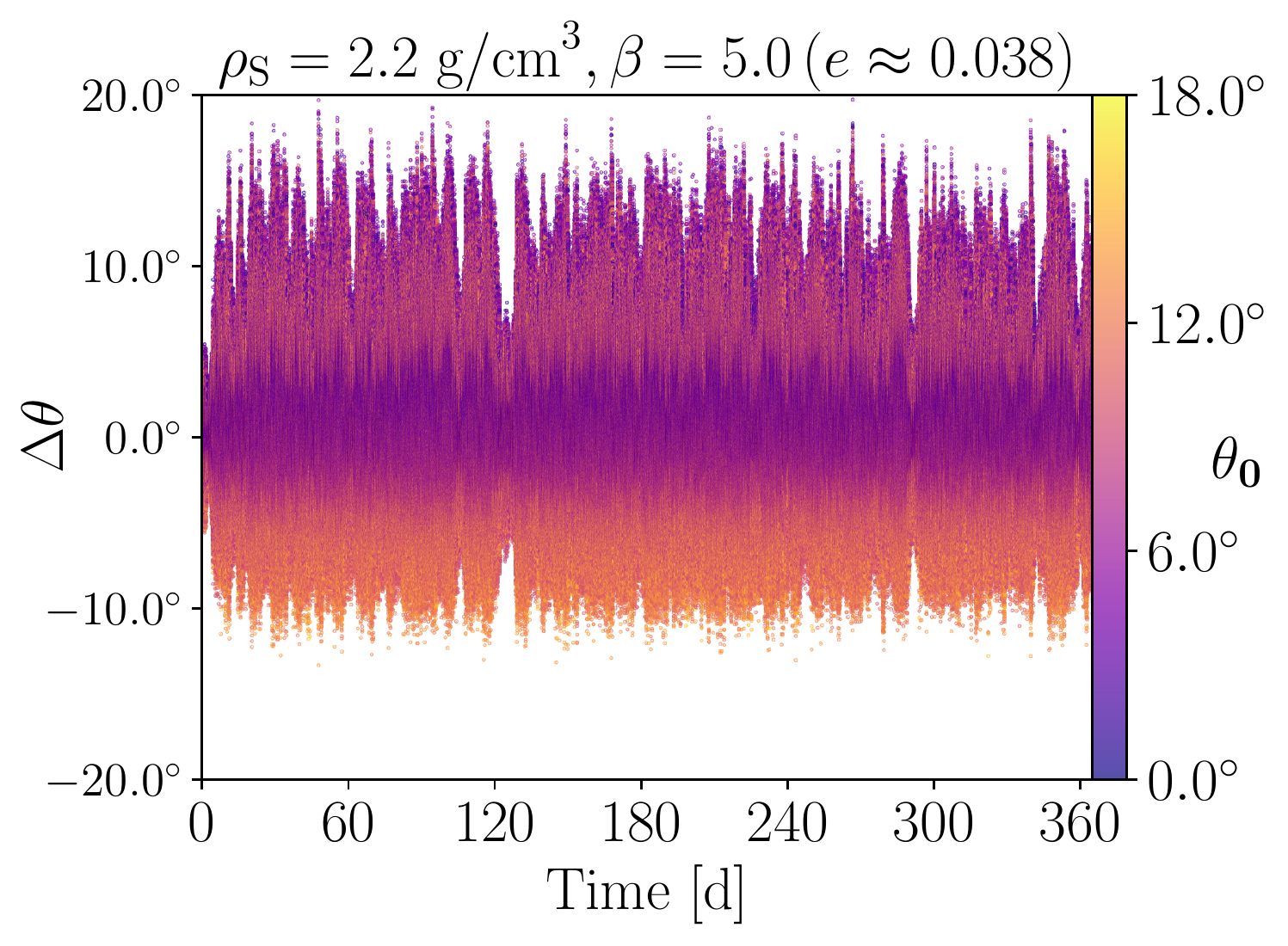}
         (b) 
         \includegraphics[clip, trim=8.5cm 0.75cm 9.5cm 0.75cm, width=\textwidth]{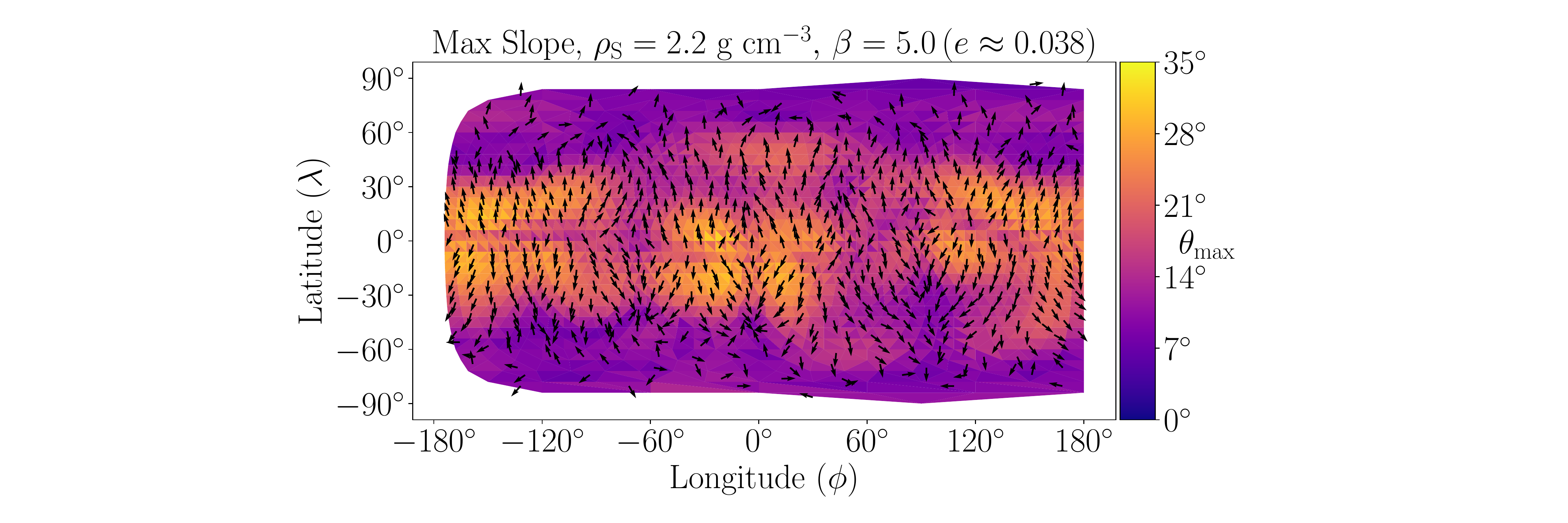}
         (c)
      \end{minipage}        
      \caption{\label{app:fig:beta5}  The spin, orbit, and surface slope evolution for $\beta=5$ and $\rho_\text{S}=2.2 \textrm{ g cm}^{-3}$ over 365 days. (a) Body separation, Euler angles, and body spin of Dimorphos. (b) Corresponding change in slope ($\Delta\theta$) over time for each surface facet, colored by the starting slope ($\theta_0$) of that facet. The surface slope evolution is dominated by NPA rotation. (c) Maximum surface slope achieved on each facet over the full 365 d simulation, with arrows pointing in the down-slope direction. }
   \end{figure*}

Here we provide supplemental plots of Dimorphos's spin and orbit, as well as its surface slope evolution, extending out to the full 365 days. The plots below are only for the nominal bulk density of $\rho_\text{S}=2.2\text{ g cm}^{-3}$. When $\beta=1$ (Fig.\ \ref{app:fig:beta1}), the orbital eccentricity remains relatively low, keeping Dimorphos in a stable rotation state, which results in small changes to the surface slopes. When $\beta$ is increased to 2 (Fig.\ \ref{app:fig:beta2}), Dimorphos becomes attitude unstable. Due to increased NPA rotation, we see much larger changes to the surface slopes. The slopes also vary chaotically since Dimorphos's spin state is chaotic. When $\beta=3$ (Fig.\ \ref{app:fig:beta3}), Dimorphos is not only in NPA rotation, but it also enters the barrel instability, characterized by rotation about its long axis. As $\beta$ increases further (Figs.\ \ref{app:fig:beta4} and \ref{app:fig:beta5}), Dimorphos's spin and orbit are increasingly perturbed, leading to larger tidal and rotational accelerations that result in larger changes in the surface slopes.

\end{appendix}

\end{document}